\newif\ifdraft
\draftfalse
\ifdraft
\documentclass[manuscript]{aastex62}
\else
\documentclass{emulateapj}
\fi

\usepackage{amsmath,here,afterpage,color,multirow,lineno}

\received{}
\revised{}
\accepted{}
\shorttitle{ALMA Images of the Host Cloud of the Intermediate-mass Black Hole Candidate CO-0.40-0.22$^*$}
\shortauthors{}
\tighten

\renewcommand\textcolor[2]{#2}
\newcommand\remove[1]{}

\usepackage{float}

\newcommand\pcc{\mathrm{cm^{-3}}}

\newcommand\kelvin{\ifmmode\mathrm{ K}\else K\fi}

\newcommand\kmps{\ifmmode\mathrm{km\,s^{-1}}\else$\mathrm{km\,s^{-1}}$\fi}

\newcommand\pc{\mathrm{pc}}

\newcommand\yr{\mathrm{yr}}

\newcommand\Msol{\ifmmode{M_\odot}\else${M_\odot}$\fi}

\newcommand\Tkin{\ifmmode{T_{\rm kin}}\else{$T_{\mathrm{kin}}$}\fi}
\newcommand\Td{\ifmmode{T_{\mathrm{d}}\else{$T_{\mathrm{d}}$}\fi}}

\newcommand\Trot{\ifmmode{T_{\rm rot}}\else{$T_{\rm rot}$}\fi}
\newcommand\Tsys{\ifmmode{T_{\rm sys}}\else{$T_{\rm sys}$}\fi}

\newcommand\nH{\ifmmode{n_{\rm H}}\else${n_{\rm H}}$\fi}
\newcommand\nHH{\ifmmode{n_{\rm H_2}}\else{$n_{\rm H_2}$}\fi}
\newcommand\NHH{\ifmmode{N_{\rm H_2}}\else{$N_{\rm H_2}$}\fi}
\newcommand\NHHavg{\ifmmode{\left\langle N_{\rm H_2}\right\rangle}\else{$\left\langle N_{\rm H_2}\right\rangle$}\fi}
\newcommand\ncrit{\ifmmode{n_{\rm crit}}\else{$n_{\rm crit}$}\fi}

\newcommand\vlsr{\ifmmode{v_{\rm LSR}}\else${v_{\rm LSR}}$\fi}
\newcommand\Mvt{\ifmmode{M_{\rm VT}}\else${M_{\rm VT}}$\fi}
\newcommand\Tmb{\ifmmode{T_{\rm MB}}\else{$T_{\mathrm{MB}}$}\fi}

\newcommand\vcol{\ifmmode{v_{\rm col}}\else{$v_{\rm col}$}\fi}

\newcommand\HOCp{\ifmmode{\rm HOC^+}\else${\mathrm HOC^+}$\fi}

\newcommand\HHHp{\ifmmode{\rm H_3^+}\else{${\rm H_3^+}$}\fi}

\newcommand\HCN{\ifmmode{\rm HCN}\else{HCN}\fi}
\newcommand\HCOp{\ifmmode{\rm HCO^+}\else{$\mathrm{HCO^+}$}\fi}
\newcommand\HCNt{\ifmmode{\rm H{^{13}C}N}\else{$\mathrm{H{^{13}C}N}$}\fi}
\newcommand\HNC{\ifmmode{\rm HNC}\else{$\mathrm{HNC}$}\fi}
\newcommand\HCCCN{\ifmmode{\rm HC_3N}\else{$\mathrm{HC_3N}$}\fi}
\newcommand\HCOpt{\ifmmode{\rm H^{13}CO^+}\else{$\mathrm{H^{13}CO^+}$}\fi}

\newcommand\Cn{\ifmmode {\rm C^0}\else $\mathrm{C^0}$\fi}

\newcommand\COt{\ifmmode{\rm {^{13}CO}}\else{$\mathrm{^{13}CO}$}\fi}
\newcommand\Ct{\ifmmode{\rm {^{13}C}}\else{$\mathrm{^{13}C}$}\fi}

\newcommand\NNHp{\ifmmode{\rm N_2H^+}\else{$\mathrm{N_2H^+}$}\fi}
\newcommand\HHCS{\ifmmode{\rm H_2CS}\else{$\mathrm{H_2CS}$}\fi}
\newcommand\CtS{\ifmmode{\rm ^{13}CS}\else{$\mathrm{^{13}CS}$}\fi}
\newcommand\OCS{\ifmmode{\rm OCS}\else{$\mathrm{OCS}$}\fi}
\newcommand\methanol{\ifmmode{\rm CH_3OH}\else{$\mathrm{CH_3OH}$}\fi}
\newcommand\methanimine{\ifmmode{\rm CH_2NH}\else{$\mathrm{CH_2NH}$}\fi}
\newcommand\pformaldehyde{\ifmmode{p\mbox{-}\rm H_2CO}\else{$p$-$\mathrm{H_2CO}$}\fi}
\newcommand\formaldehyde{\ifmmode{\rm H_2CO}\else{$\mathrm{H_2CO}$}\fi}
\newcommand\pfx{\ifmmode{3_{21}--2_{20}}\else{$3_{21}--2_{20}$}\fi}
\newcommand\pfy{\ifmmode{3_{22}--2_{21}}\else{$3_{22}--2_{21}$}\fi}
\newcommand\HHO{\ifmmode{\rm H_2O}\else{$\mathrm{H_2O}$}\fi}

\newcommand\JJ[2]{\ifmmode{\mbox{{\it J}={#1}\mbox{--}{#2}}}\else{{\it J}={#1}--{#2}}\fi}
\newcommand\JK[4]{\ifmmode{{J_K}=#1_{#2}\mbox{--}#3_{#4}}\else{${\it J_K}=#1_{#2}\mbox{--}#3_{#4}$}\fi}
\newcommand\xJJ[2]{\ifmmode{\mbox{{#1}\mbox{--}{#2}}}\else{{#1}--{#2}}\fi}
\newcommand\xJK[4]{\ifmmode{#1_{#2}\mbox{--}#3_{#4}}\else{$#1_{#2}\mbox{--}#3_{#4}$}\fi}

\newcommand\CIa{\ifmmode{^3}P_1\mbox{--}{^3}P_0\else${^3}P_1\mbox{--}{^3}P_0$\fi}

\newcommand\gl{\ifmmode l\else{\it l}\fi}
\newcommand\gb{\ifmmode b\else{\it b}\fi}

\newcommand\CLb{CO$-0.30$$-0.07$}
\newcommand\theObj\CLb

\newcommand\Dv{\ifmmode{\Delta v}\else{$\Delta v$}\fi}
\newcommand\vc{\ifmmode{\left<v\right>}\else{$\left<v\right>$}\fi}

\newcommand{\avir}{\ifmmode{\alpha_{\rm vir}}\else{$\alpha_{\rm vir}$}\fi}

\newcommand{\RN}[1]{\textup{\uppercase\expandafter{\romannumeral#1}}}
\newcommand\Sdust{\ifmmode{S_{500}}\else{$S_{500}$}\fi}
\newcommand\II[1]{\ifmmode{I_{#1}}\else{$I_{#1}$}\fi}

\newcommand\myvector[1]{\ifmmode{\mbox{\boldmath ${#1}$}}\else{\boldmath {${#1}$}}\fi}
\newcommand{\Rt}{\ifmmode{{R_{13}}}\else${R_{13}}$\fi}
\newcommand{\Iobs}{\ifmmode{I}\else${I}$\fi}
\newcommand{\Icalc}{\ifmmode{{F}\left(\pv\right)}\else${{F}\left(\pv\right)}$\fi}
\newcommand{\Icalci}{\ifmmode{{F}\left(\pv^{(j)}\right)}\else${{F}\left(\pv^{(j)}\right)}$\fi}
\newcommand{\Icalcv}{\ifmmode{\myvector{F}\left(\pv\right)}\else${\myvector{F}\left(\pv\right)}$\fi}
\newcommand{\xmol}{\ifmmode{{x_{\rm mol}}}\else${x_{\rm mol}}$\fi}
\newcommand{\xmolp}[1]{\ifmmode{{x_{\rm mol}\left(#1\right)}}\else${x_{\rm mol}\left(\mbox{#1}\right)}$\fi}
\newcommand{\ff}{\ifmmode{\Phi}\else${\Phi}$\fi}
\newcommand{\fff}{\ifmmode{\phi}\else${\phi}$\fi}
\newcommand{\fcal}{\ifmmode{f_{\rm cal}}\else${f_{\rm cal}}$\fi}

\newcommand{\Ea}{\ifmmode\epsilon^{\rm a}\else$\epsilon_{\rm a}$\fi}
\newcommand{\Em}{\ifmmode\epsilon\else$\epsilon$\fi}

\newcommand{\sa}{\ifmmode\sigma\else$\sigma$\fi}
\newcommand{\sm}{\ifmmode\sigma^{\rm m}\else$\sigma_{\rm m}$\fi}
\newcommand{\scal}{\ifmmode\sigma_{\rm cal}\else$\sigma_{\rm cal}$\fi}

\newcommand{\pv}{\myvector{p}}

\newcommand{\PDF}[1]{{\ifmmode P(#1) \else $P(#1)$ \fi}}

\newcommand{\NN}[1]{\ifmmode N\left(#1\right)\else$N\left(#1\right)$\fi}
\newcommand\RRx{\ifmmode R_{43}\else$R_{43}$\fi}
\newcommand\IIx{\ifmmode I_{13}\else$I_{13}$\fi}
\newcommand\Ihnc{\ifmmode I_{\hnc}\else$I_{13}$\fi}
\newcommand\dNdv{\ifmmode {\mathrm{d}N}/{\mathrm{d}v}\else${\mathrm{d}N}/{\mathrm{d}v}$\fi}
\newcommand\dNHdv{\ifmmode \frac{\mathrm{d}N_{\rm H_2}}{\mathrm{d}v}\else$\frac{\mathrm{d}N_{\rm H_2}}{\mathrm{d}v}$\fi}
\newcommand\Xdvdr{\ifmmode {{X}/{\frac{\mathrm{d}v}{\mathrm{d}r}}}\else{${X}/\frac{\mathrm{d}v}{\mathrm{d}r}$}\fi}

\newcommand\Nobs{\ifmmode{N_{\rm obs}}\else{$N_{\rm obs}$}\fi}
\newcommand\Np{\ifmmode{N_{p}}\else{$N_{p}$}\fi}

\newcommand\Mmag{\ifmmode M_\Phi\else$M_\Phi$\fi}
\newcommand\nth{\ifmmode {n_{\mathrm{th}}}\else$n_{\mathrm{th}}$\fi}
\newcommand\SFRff{\ifmmode {\mathrm{SFR_{ff}}}\else$\mathrm{SFR_{ff}}$\fi}

\newcommand\zCR{\ifmmode \zeta_{\mathrm{CR}}\else$\zeta_{\mathrm{CR}}$\fi}
\newcommand\xe{\ifmmode x_{\mathrm{e}}\else$x_{\mathrm{e}}$\fi}

\renewcommand\theObj{\ifmmode{\mathrm{CO{-0.4}}}\else{CO--0.4}\fi}

\ifdraft
\pagewiselinenumbers
\fi

\begin{document}

\title{ALMA Images of the Host Cloud of the Intermediate-mass Black Hole Candidate CO-0.40-0.22$^*$:  No Evidence for Cloud--Black Hole Interaction, but Evidence for a Cloud-Cloud Collision}
\author{Kunihiko Tanaka}
\email{ktanaka@phys.keio.ac.jp}
\affil{Department of Physics, Faculty of Science and Technology, Keio University, 3-14-1 Hiyoshi, Yokohama, Kanagawa 223--8522 Japan}



\begin{abstract}
This paper reports a re-analysis of archival ALMA data of the high velocity (-width) compact cloud (HVCC) CO${-0.40}{-0.22}$, which has recently been hypothesized to host an intermediate-mass black Hole (IMBH).
If beam-smearing effects, difference in beam sizes among frequency bands, and \textcolor{red}{Doppler shift due to the motion of the Earth} are considered accurately, none of the features reported as evidence for an IMBH in previous studies are confirmed in the re-analyzed ALMA images.
Instead, through analysis of the position--velocity structure of the HCN $J$=$3$--$2$ data cube, we have found kinematics typical of a cloud--cloud collision (CCC), namely, two distinct velocity components bridged by broad emission features with elevated temperatures and/or densities.  One velocity component has a straight filamentary shape with approximately constant centroid velocities along its length but with a steep, V-shaped velocity gradient across its width.
This contradicts the IMBH scenario but is consistent with a collision between two dissimilar-sized clouds. 
From a non-LTE analysis of the multi-transition methanol lines, the volume density of the post-shock gas has been measured to be $\gtrsim 10^6\ \pcc$, indicating that the CCC shock can compress gas  in a short timescale to densities typical of star-forming regions.
Evidence for star formation has not been found, possibly because the cloud is in an early phase of CCC-triggered star formation or because the collision is non-productive.
\end{abstract}

\section{INTRODUCTION}
High velocity(-width) compact clouds (HVCCs) are a population of dense gas clumps with peculiarly broad-lined molecular emissions having full-width-zero-intensity (FWZI) width of a few tens to 100 \kmps\ \citep{Oka2012,Tanaka2015}.
Approximately 100 HVCCs and similar broad molecular line features have been identified in the central molecular zone (CMZ) in the Galaxy \textcolor{red}{\citep{Oka2012,Tokuyama2017}}.
They are approximately a factor of 5 above the standard size--linewidth relation for the CMZ clouds, indicating of the ubiquitous presence of energy-injecting events in the CMZ.
Supernova (SN)--cloud interactions \citep{Tanaka2009,Yalinewich2017} and cloud--cloud collisions \citep[CCCs;][]{Tanaka2015,Ravi2017} have been proposed as candidates for these energy sources.

Recently, following a series of observational studies, \cite{Oka2016,Oka2017} have proposed a new hypothesis for the formation of the archetypical HVCC CO${-0.40}{-0.22}$ (hereafter \theObj): gravitational acceleration by an intermediate-mass black hole (IMBH).
The cloud \theObj\ is one of the most energetic HVCCs, which have $90\ \kmps$-wide HCN and SiO emissions \citep{Tanaka2015,Oka2016}. 
\cite{Oka2016} have interpreted this broad-lined molecular emission as a high velocity gas stream trailing from a clump that was kicked by the gravitational interaction with an IMBH of $10^5\ \Msol$.
Although this clump--stream structure was not fully resolved in their original single-dish images, a follow-up interferometric study using the Atacama Large Millimeter--submillimeter Array (ALMA) has identified a clump with extremely broad-lined emission, $110\ \kmps$-wide FWZI, at the position of the gravitationally kicked clump predicted by their model \citep{Oka2017}.
In addition, they claim that a point-like continuum source near the clump, \theObj$^*$, is self-absorbed synchrotron emission from an IMBH, based on the shallow spectral index $\alpha = 1.18\pm0.65$  measured in the 230--265 GHz frequency range.

This ``gravitational-kick'' formation scenario seems to explain successfully the cloud kinematics, but it should be noted that the model is based on single-dish images with insufficient resolution. 
Although a candidate for the gravitationally kicked clump has been identified in the high-resolution ALMA images, a direct comparison between the gas stream model and the position--velocity (PV) structure of the high-resolution ALMA image has yet to be performed.
In addition, the significance of the non-thermal spectral energy distribution (SED) ascribed to the IMBH candidate has been called into question by a recent follow-up study.
\cite{Ravi2017} reported a non-detection of $\theObj^*$ at 34.25 GHz, which is contradictory to the shallow value of $\alpha$ obtained from the ALMA data and instead indicates thermal emission from warm dust.
\textcolor{red}{Furthermore, it is not clear whether such exotic objects as IMBHs could create the HVCCs which are commonly found in the CMZ, as questioned by \cite{Yalinewich2017}; 
these authors argue that the sum of the cloud mass and IMBH mass must be greater than the total mass of the CMZ in order to explain the observed HVCC formation rate with the IMBH scenario.}

In this present paper, we re-examine the gravitational kick scenario by analyzing the same ALMA data used in \cite{Oka2017}. 
We find no evidence for gravitational interaction with an IMBH in the re-analyzed ALMA data, when beam-smearing effect and \textcolor{red}{time variation of the Doppler-shifted frequencies due to the motion of the Earth}  are considered accurately.
We investigate a CCC hypothesis as an alternative for the origin of the broad-lined molecular emission from \theObj.

\section{Re-analysis of the ALMA Archival Data }
We re-analyzed the ALMA data used by \cite{Oka2017}, which we obtained from the ALMA science archive (id: 2012.1.00940.S).  
The data include multiple datasets observed with different antenna-array configurations from July 2013 to January 2015.
Table \ref{TABLE_DATASETS} summarizes information about the datasets, with the observations labeled by the execution dates. 
The spectral data were taken in three frequency windows centered at 230 GHz,  250 GHz, and 265 GHz.  
The 250 GHz and 265 GHz data were obtained simultaneously.     
The 230- and 265-GHz bands have a $\sim 900$ MHz bandwidth and a $\sim 200\ $kHz channel separation, and observed with both the 12-m and 7-m arrays.     
The 250-GHz data was taken with a broad band mode with a $\sim 2$ GHz bandwidth and a 15 MHz channel width, and lacks 7-m array data. 

We employed standard data reduction procedures, using the Common Astronomy Software Applications (CASA) package developed by the National Radio Astronomy Observatory (NRAO);  flagging low quality data, correction for atmospheric conditions, and calibration
of the visibility data were performed using the reduction script provided by the ALMA observatory.  
The calibrators used for bandpass, flux, and complex gain calibrations are listed in Table \ref{TABLE_DATASETS}.   
The {\tt 08JUL2013} data lacks calibration data for the system noise temperature for multiple antennas; therefore, for this dataset we substituted the averaged values for the other antennas.

The main target lines are CO \JJ{2}{1} (230.538 GHz) and HCN \JJ{3}{2} (265.886 GHz).
We also detected more than 20 other lines, which are listed in Table \ref{TABLE_LINES}, in the three spectral windows.  
The spectra in the 230- and 265-GHz bands are separated into line and continuum components using the {\tt UVCONTSUB} task in the CASA package. 
Continuum subtraction was not possible for the 250-GHz band data due to the extreme crowdedness of the emission lines.

We performed mosaic imaging using the {\tt TCLEAN} task.
Briggs weighting with a robustness parameter of 1 was applied for the 230-GHz band line data, and the natural weighting was applied for the 250- and 265-GHz line data.  
For the continuum maps, robustness of $1$  and a $uv$-taper at the $1''$ angular scale were applied for the 230 GHz and 265 GHz data, respectively, to improve the signal-to-noise ratio.
The synthesized beam sizes and beam position angles are listed in Table \ref{TABLE_MAPS}.   
The 250 GHz and 265 GHz maps have smaller beam widths than the 230 GHz maps, as the former were obtained with more extended antenna configurations.
We refined the image quality by performing self-calibration cycles using the moment-0 CO and HCN images independently for each dataset.
We excluded the {\tt 08JUL2013} data from the final imaging, since the moment-0 map constructed from this data was inconsistent with the images from other 265-GHz band data, presumably due to incorrect calibration and the lack of correct \Tsys\ measurement data.

\ifdraft
\begin{deluxetable}{lcccc}
\tablecaption{ALMA Data\label{TABLE_DATASETS}}


\tablecolumns{5}
\small
\tablehead{\\
\colhead{Dataset\tablenotemark{a}} & \colhead{array} & \colhead{band} & \colhead{Calibrators\tablenotemark{b}}  & \colhead{$uv$\tablenotemark{c}}\\
             &                 & \colhead{GHz}  & \colhead{flux/bandpass/gain} & \colhead{k$\lambda$} 
}
\tabletypesize{\scriptsize}
\tablewidth{0pt}
\startdata
\texttt{07OCT2013} & 7-m  & 230 & Nep./J1924/J1744& 7--37.5\\
\texttt{01NOV2013} & 7-m  & 230 & Nep./J1924/J1744& 5.5--37.5 \\
\texttt{22JAN2015} & 12-m & 230 & Titan/J1517/J1744 & 10--252  \\
\texttt{13APR2014} & 7-m  & 265 & Mar./J1427/J1744 & 7--42\\
\texttt{29DEC2014} & 7-m  & 265 & Titan/J1733/J1744 & 8.5--40 \\
\texttt{08JUL2013}\tablenotemark{d} & 12-m & 250,265  &J2232/J2258/J2337 & 10--390 \\
\texttt{11MAR2014} & 12-m & 250,265  & Titan/J1517/J1745& 11.5--332 
\enddata
\tablenotetext{a}{Named according to the execution dates; {\tt 07OCT2013} was obtained on October 7, 2013, and so on. }
\tablenotetext{b}{Abbreviations for the calibrator source names: Nep. = Neptune, Mar. = Mars, J1924 = J1924-2924,  J1744 = J1744-3116, J1517 = J1517-2422, J1427 = J1427-4206, J1733 = J1733-1304, J2322 = J2322+117,  J2258 = J2258-2758,  J2337 = J2337-230,  J1745 = J1745-290}
\tablenotetext{c}{Values after flagging.}
\tablenotetext{d}{Lacks correct $\Tsys$ calibration. Not used in the final images.}


\end{deluxetable}
\else
\begin{deluxetable}{lcccc}
\tablecaption{ALMA Data\label{TABLE_DATASETS}}


\tablecolumns{5}
\small
\tablehead{\\
\colhead{Dataset\tablenotemark{a}} & \colhead{array} & \colhead{band} & \colhead{Calibrators\tablenotemark{b}}  & \colhead{$uv$\tablenotemark{c}}\\
             &                 & \colhead{GHz}  & \colhead{flux/bandpass/gain} & \colhead{k$\lambda$} 
}
\tabletypesize{\scriptsize}
\tablewidth{0pt}
\startdata
\texttt{07OCT2013} & 7-m  & 230 & Nep./J1924/J1744& 7--37.5\\
\texttt{01NOV2013} & 7-m  & 230 & Nep./J1924/J1744& 5.5--37.5 \\
\texttt{22JAN2015} & 12-m & 230 & Titan/J1517/J1744 & 10--252  \\
\texttt{13APR2014} & 7-m  & 265 & Mar./J1427/J1744 & 7--42\\
\texttt{29DEC2014} & 7-m  & 265 & Titan/J1733/J1744 & 8.5--40 \\
\texttt{08JUL2013}\tablenotemark{d} & 12-m & 250,265  &J2232/J2258/J2337 & 10--390 \\
\texttt{11MAR2014} & 12-m & 250,265  & Titan/J1517/J1745& 11.5--332 
\enddata
\tablenotetext{a}{Named according to the execution dates; {\tt 07OCT2013} was obtained on October 7, 2013, and so on. }
\tablenotetext{b}{Abbreviations for the calibrator source names: Nep. = Neptune, Mar. = Mars, J1924 = J1924-2924,  J1744 = J1744-3116, J1517 = J1517-2422, J1427 = J1427-4206, J1733 = J1733-1304, J2322 = J2322+117,  J2258 = J2258-2758,  J2337 = J2337-230,  J1745 = J1745-290}
\tablenotetext{c}{Values after flagging.}
\tablenotetext{d}{Lacks correct $\Tsys$ calibration. Not used in the final images.}


\end{deluxetable}
\fi

\ifdraft
\begin{deluxetable}{lcrl}
\tablecaption{Detected Lines\label{TABLE_LINES}}


\tablecolumns{4}
\small
\tablehead{\\ \colhead{molecule} & \colhead{transition} & \colhead{frequency} & notes \\
  & & \colhead{(GHz)} &   }
\tabletypesize{\scriptsize}
\tablewidth{0pt}
\startdata
\methanol         & \xJK{3}{-2}{4}{-1}E        & 230.027 \\
CO                & \xJJ{2}{1}                 & 230.538 \\
\methanimine      & \xJK{7}{1,6}{7}{0,7}       & 250.162 \\
\methanol         & \xJK{13}{3,10}{13}{2,11}A$^{-+}$  & 250.291 \\
NO                & \xJJ{5/2}{3/2}             & 250.44 \\
                  &                           & 250.45  \\
\methanol         & \xJK{11}{0,11}{10}{1,10}A$^{++}$  & 250.507 \\
NO                & \xJJ{5/2}{3/2}             & 250.708  \\
                  &                           & 250.816  \\
\methanol         & \xJK{12}{3,9}{12}{2,10}A$^{-+}$      & 250.635 \\
                  & \xJK{11}{3,8}{11}{2,9}A$^{-+}$    & 250.924\\
$\mathrm{SO_2}$   & \xJK{13}{1,13}{12}{0,12}   & 251.200  &  \multirow{2}{*}{$\left.\right\}$ blended} \\
                  & \xJK{8}{3,5}{8}{2,6}       & 251.211  &  \\               
c-$\mathrm{HCCCH}$& \xJK{6}{2,5}{5}{1,4}       & 251.314 \\
\methanol         & \xJK{9}{3,6}{9}{2,7}A$^{-+}$      & 251.360 \\
\methanimine      & \xJK{6}{0,6}{5}{1,5}       & 251.421 \\
\methanol         & \xJK{8}{3,5}{8}{2,6}A$^{-+}$      & 251.517 &  \multirow{2}{*}{$\left.\right\}$ blended} \\
c-$\mathrm{HCCCH}$& \xJK{7}{1,7}{6}{0,6}       & 251.527 & \\
t-$\mathrm{CH_3CH_2OH}$ & \xJK{15}{1,15}{14}{0,14} & 251.567  & \\
\methanol         & \xJK{7}{3,4}{7}{2,5}A$^{-+}$      & 251.642 \\
                  & \xJK{6}{3,3}{6}{2,4}A$^{-+}$      & 251.739 \\
                  & \xJK{5}{3,2}{5}{2,3}A$^{-+}$      & 251.812 \\
\methanimine      & \xJK{4}{1,3}{3}{1,2}       & 266.270 \\
HCN               & \xJJ{3}{2}                 & 265.886 
\enddata

\end{deluxetable}
\else
\begin{deluxetable}{lcrl}
\tablecaption{Detected Lines\label{TABLE_LINES}}

\end{deluxetable}
\fi

\ifdraft
\begin{deluxetable}{lcccc}
\tablecaption{Image Data\label{TABLE_MAPS}}


\tablecolumns{5}
\small
\tablehead{\\ \colhead{band} & \colhead{line/continuum} & \colhead{array} & \colhead{$b_{\rm maj} \times b_{\rm min}$} & \colhead{beam P.A.} \\
            \colhead{GHz} &                          &                   & \colhead{arcsec$^2$}          & \colhead{deg.} }
\tabletypesize{\scriptsize}
\tablewidth{0pt}
\startdata
230 & continuum & 12+7 & $2.03\times1.24$ & 74.6 \\
265 & continuum & 12+7 & $1.98\times1.35$ & 71.9 \\
230 & line      & 12+7 & $1.75\times1.02$ & 74.9 \\
265 & line      & 12+7 & $1.35\times0.60$ & 69.8 \\
250 & line+continuum  & 12 & $1.42\times0.59$ & 69.9 
\enddata

\end{deluxetable}
\else
\begin{deluxetable}{lcccc}
\tablecaption{Image Data\label{TABLE_MAPS}}


\tablecolumns{5}
\small
\tablehead{\\ \colhead{band} & \colhead{line/continuum} & \colhead{array} & \colhead{$b_{\rm maj} \times b_{\rm min}$} & \colhead{beam P.A.} \\
            \colhead{GHz} &                          &                   & \colhead{arcsec$^2$}          & \colhead{deg.} }
\tabletypesize{\scriptsize}
\tablewidth{0pt}
\startdata
230 & continuum & 12+7 & $2.03\times1.24$ & 74.6 \\
265 & continuum & 12+7 & $1.98\times1.35$ & 71.9 \\
230 & line      & 12+7 & $1.75\times1.02$ & 74.9 \\
265 & line      & 12+7 & $1.35\times0.60$ & 69.8 \\
250 & line+continuum  & 12 & $1.42\times0.59$ & 69.9 
\enddata

\end{deluxetable}
\fi

\section{RESULTS}

\ifdraft
\begin{figure*}[p!]
\else
\begin{figure*}[h!]
\fi
\epsscale{1}
\plotone{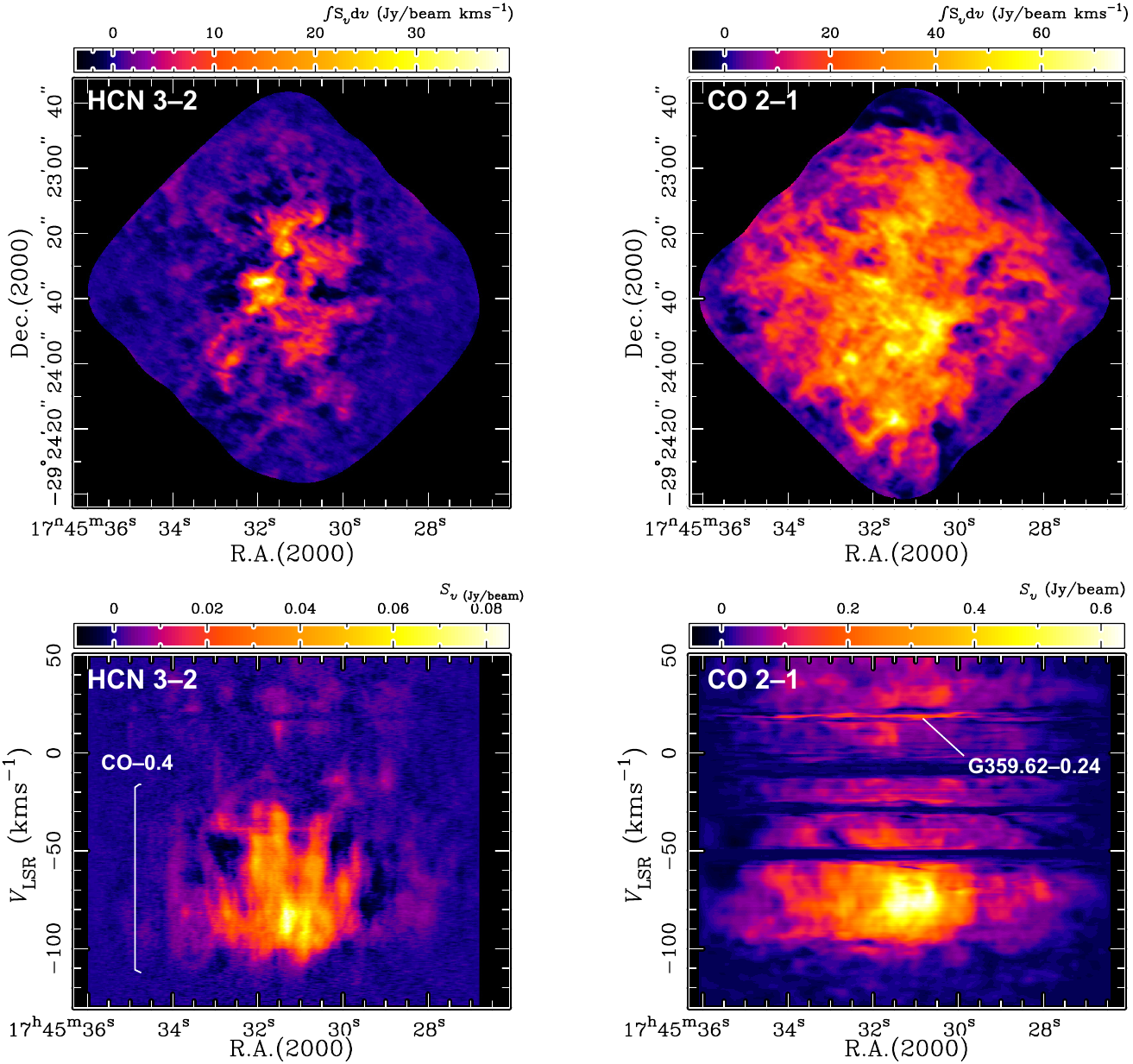}
\caption{(top panels) HCN \JJ{3}{2}\ and CO \JJ{2}{1}\ flux densities integrated over the velocity range $-110\ \kmps$ to $0$\ \kmps.  (bottom panels) Position-velocity diagrams for  HCN \JJ{3}{2}\ and CO \JJ{2}{1}\ projected on the right ascension (R.A.) axis, averaged over the full declination (Dec.) range.   The emission from the \theObj\ exists primarily in the velocity range $-110$ to $-10\ \kmps$.  Clouds with  positive velocities are the foreground cloud G359.62${-0.24}$ or Galactic center clouds unrelated to \theObj.\label{FIG1}}
\end{figure*}

\ifdraft
\begin{figure*}[p!]
\else
\begin{figure*}[t!]
\fi
\epsscale{1}
\plotone{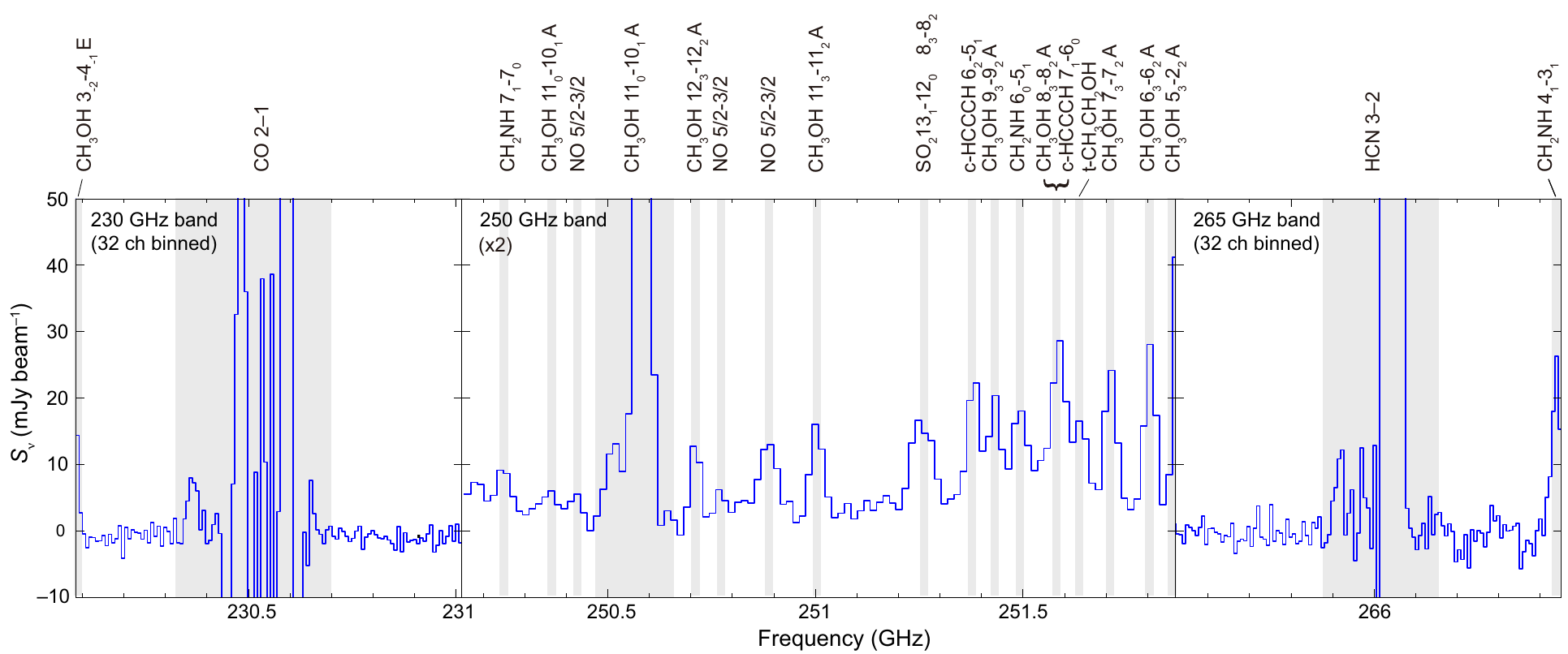}
\caption{Spectra of the 230, 250, and 265 GHz bands averaged over the 5 intensity peaks indicated in Figure \ref{FIG_FILAMENT}.   The 230 and 265 GHz spectra are averaged over every 32 channels.\label{FIG_SPECALL}}
\end{figure*}

\ifdraft
\begin{figure*}[p!]
\else
\begin{figure*}[t!]
\fi
\epsscale{0.9}
\plotone{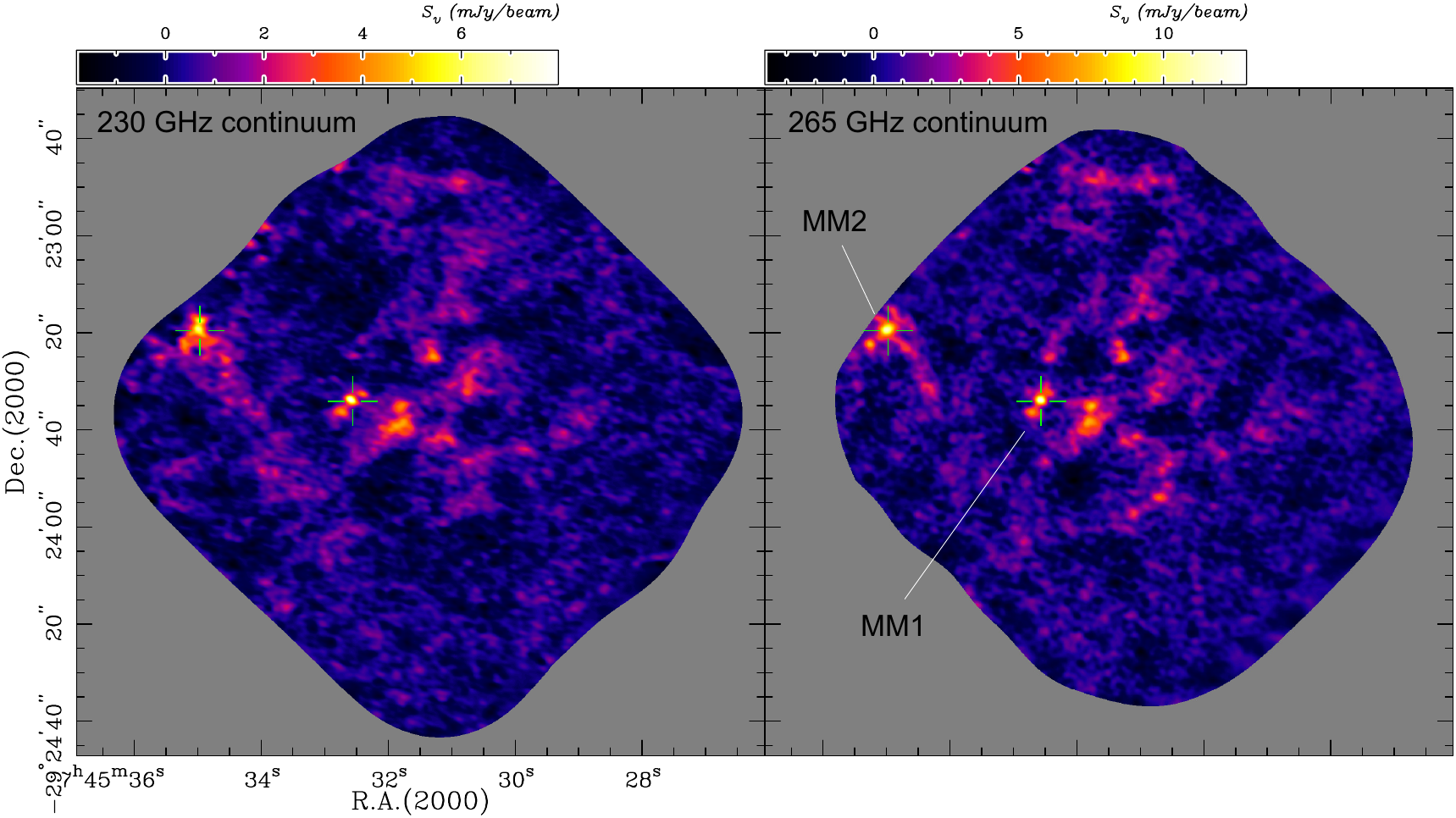}
\caption{Continuum maps of the 230-GHz and 265-GHz bands.  Positions of the two bright point-like sources, MM1 and MM2, are indicated by cross marks.\label{FIG2}}
\plotone{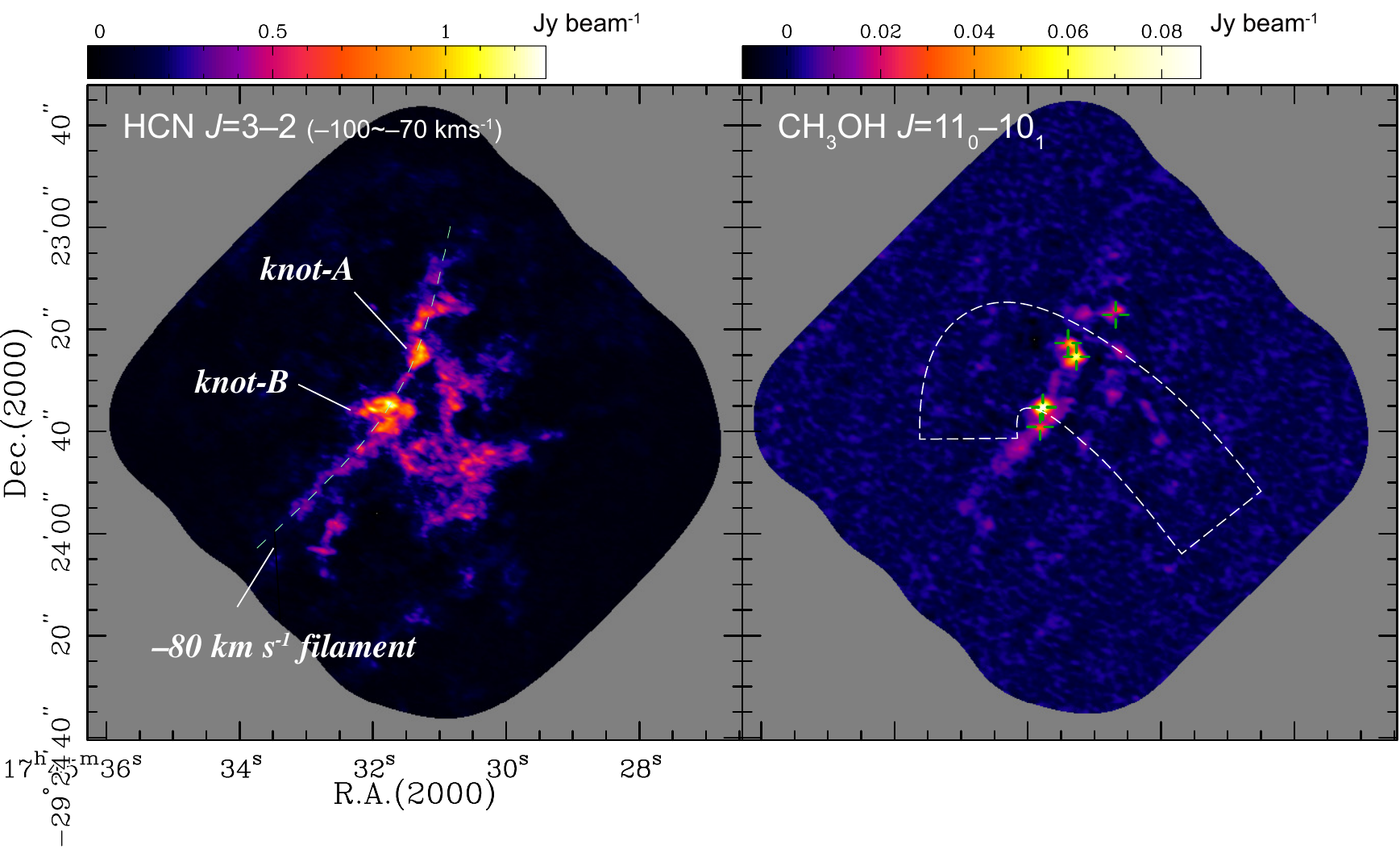}
\caption{Peak intensity maps for HCN \JJ{3}{2} (left) and \methanol $11_0$--$10_1$ A$^{++}$ (right), calculated from data smoothed into $2\ \kmps$ and $36\ \kmps$ velocity bins, respectively.   The HCN \JJ{3}{2}\ intensities are measured for the velocity range $-100$ to $-70$\ \kmps.  The shape of the $-80$-\kmps\ filament is indicated in the HCN map, with two prominent features, labeled knot-A and knot-B.  The crosses on the \methanol\ map are the positions of the averaged spectra used in Figure \ref{FIG_SPECALL} and the population diagram analysis (\S\ref{ROT_ANALY}).   An approximate shape for the hypothesized gas stream (read from Figure 3 of \cite{Oka2016}) is indicated by the dashed lines.\label{FIG_FILAMENT}}
\end{figure*}

\ifdraft
\begin{figure}[p!]
\else
\begin{figure}[t!]
\fi
\ifdraft
\epsscale{.6}
\else
\epsscale{1.}
\fi
\plotone{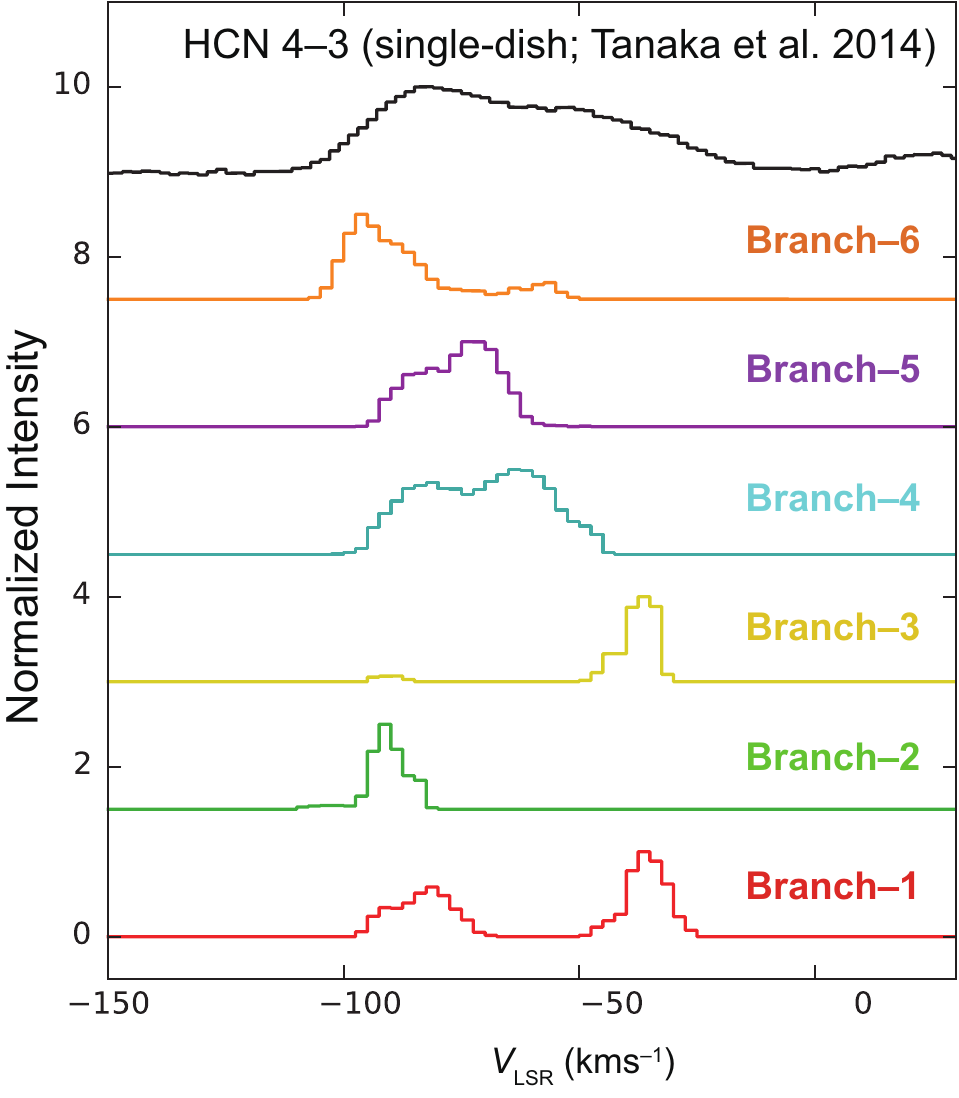}
\textcolor{red}{\caption{Spatially averaged HCN \JJ{3}{2}\ spectra of the main branches, which are normalized so that the peak intensities are unity.  The top spectrum is the HCN \JJ{4}{3}\ spectrum obtained with the ASTE 10-m telescope \citep{Tanaka2014}, averaged over the entire cloud.\label{FIG_DGRAM1}}}
\end{figure} 

\subsection{The $-80$-\kmps\ Filament and Knots}
Figure \ref{FIG1}\ shows the integrated intensity maps and the PV diagrams for the HCN \JJ{3}{2}\ and CO \JJ{3}{2}\ lines.  
The PV diagrams are plotted along the right-ascension (R.A.) axis and are averaged over the full declination range. 
Multiple velocity components exist along the line-of-sight toward the target cloud. 
The emission from \theObj\ extends from $-110$ \kmps\ to $-20$ \kmps\ in the velocity of the local standard of rest (\vlsr). 
The \COt\ \JJ{2}{1}\ image is self-absorbed at $\vlsr = -50\ \kmps$, $-25\ \kmps$, and $-5\ \kmps$ by foreground clouds in the Galactic disk, whereas the HCN \JJ{3}{2}\ data are almost free of self-absorption dips for the velocity range of \theObj.
The narrow line emission at $\vlsr = 20\ \kmps$ in the CO \JJ{2}{1}\ PV diagram is the foreground infrared dark cloud G359.62${-0.24}$  \citep[or ``Freccia Rossa'' cloud;][]{Bally2010,Ravi2017}, which is seen as an absorption dip in the HCN \JJ{3}{2}\ image.  
Other positive-velocity emissions are from Galactic center clouds unrelated to \theObj. 

The CO integrated intensity map illustrates the highly complicated cloud structure, which consists of many small, entangled filamentary features. 
In contrast, most of the spatially extended emissions are transparent in the HCN \JJ{3}{2}\ image, in which the central high-density portion of the cloud is captured owing to its high critical density (\ncrit). 
In the following section of this paper, we mainly use the HCN \JJ{3}{2}\ data to analyze the cloud structure.

Figure \ref{FIG2} shows the continuum maps at 230 GHz and 265 GHz.   
We identified two bright point-like sources that are present in both frequency bands, which we labeled MM1 and MM2.    
Source MM1 corresponds to the IMBH candidate $\theObj^*$ \citep{Oka2017}; it is not an isolated point source in our images, but is located in a ridge-like extended background emission $7''$ long.

Figure \ref{FIG_SPECALL} shows the spectra of the three spectral windows averaged over the 5 positions indicated in Figure \ref{FIG_FILAMENT}.
In addition to the CO and HCN lines, more than 20 lines from \methanol, \methanimine, NO, $\mathrm{SO_2}$, and c-$\mathrm{HCCCH}$ are detected. 
These emissions are mostly concentrated in a filament running through the central region of \theObj\ in an approximately southeast--northwest direction, as represented by the $\methanol$ $11_0$--$10_1$ map shown in Figure \ref{FIG_FILAMENT}.   
This filament has an approximately constant velocity of $\sim -80$ \kmps\ throughout the full length.
A corresponding filamentary structure is also seen in the HCN map in the $-90$ to $-70\ \kmps$ velocity range.  
We refer to this structure henceforth as the $-80$-\kmps\ filament.
The brightness of the filament in optically thin methanol and other rare molecular lines indicates that the filament is the densest part of \theObj.

The $-80$-\kmps\ filament has a few knots with bright methanol emissions.  The two brightest knots are labeled as knots A and B in Figure \ref{FIG2}.  
Both knots are also bright in the HCN \JJ{3}{2}\ image, and the filament is wider at knot B than at other portions along its length.
Knot B is located at the position where \cite{Oka2017} reported the gravitationally kicked cloud.

\textcolor{red}{\subsection{The Cloud Structure and Velocity Field\label{SECTION_DENDRO}}}

\subsubsection{Three Velocity Components\label{SECTION_MULTVEL}}

\ifdraft
\begin{figure*}[p!]
\else
\begin{figure*}[t!]
\fi
\ifdraft
\epsscale{.9}
\else
\epsscale{.9}
\fi
\plotone{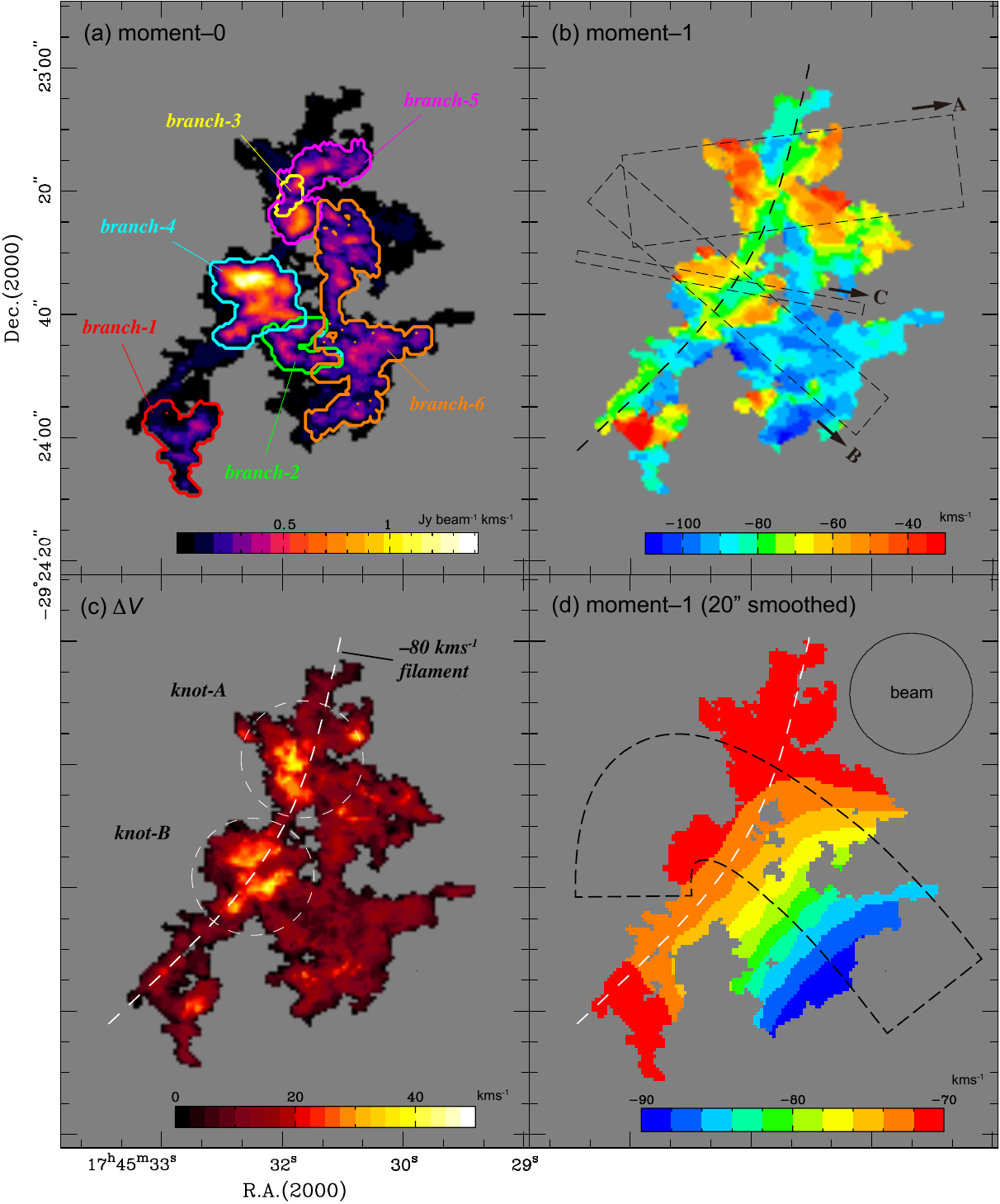}
\caption{Maps of (a) moment-0, (b) moment-1, (c) velocity widths (\Dv), and (d) moment-1 smoothed with $20''$ Gaussian for the cloud image reconstructed from the six main branches and their common trunk nodes.  The width \Dv\ is defined as the moment-1 value divided by the spectral peak intensity toward each line of sight. The 2D spatial distributions of the branches are shown by the contours overlaid on the moment-0 map.   The shape of the $-80$-\kmps\ is denoted by the dashed-lined curves on the moment-1 and \Dv\ maps, which are identical to that in Figure \ref{FIG_FILAMENT}.   The rectangles in the moment-1 map indicate the strips for the PV diagrams in Figures \ref{FIG_STRIPS} and \ref{FIG_ERROR}.  The approximate shape of the hypothesized gas stream \citep{Oka2016} is shown in panel (d).\label{FIG_DGRAM3}}
\end{figure*} 

\textcolor{red}{
In this section, we analyze the internal structure and velocity field of \theObj\ using the HCN \JJ{3}{2}\ data cube.
First we extract the primary structure of the cloud in position--position--velocity (PPV) space by means of the dendrogram method \citep{Rosolowsky2008};
this step is necessary since the HCN data cube exhibits highly complicated PPV structure, in which almost every line-of-sight contains multiple velocity components including those physically unrelated to the main body of the cloud.
We identified six main branches in the dendrogram created from the HCN data cube, and reconstructed the image of the primary PPV structure by pruning other minor structures and components unrelated to the cloud.
The details of the dendrogram analysis are described in Appendix A. 
}

\textcolor{red}{
Figure \ref{FIG_DGRAM1} shows the averaged spectra for the six main branches.
}
The figure indicates that the cloud consists of three major velocity components:  two narrow line components at $-40\ \kmps$ and $-80\ \kmps$, and a moderately broad component at $-60$\ \kmps.   
The FWZI velocity widths are 10 \kmps\ for the $-40$- and $-80$-\kmps\ components, and $50$ \kmps\ for the $-60$-\kmps\ component.
The single-dish HCN \JJ{4}{3}\ spectrum averaged over the entire cloud \citep{Tanaka2014}\ is also presented in the figure for comparison, showing that the broad profile of the averaged spectrum is the superposition of the three velocity components at $-40$, $-60$, and $-80$\ \kmps, with $\lesssim 50\ \kmps$ widths, but none of the six main branches alone can account for the $\sim90\ \kmps$ velocity width of the averaged spectrum.

\ifdraft
\begin{figure*}[p!]
\else
\begin{figure*}[t!]
\fi
\epsscale{0.8}
\plotone{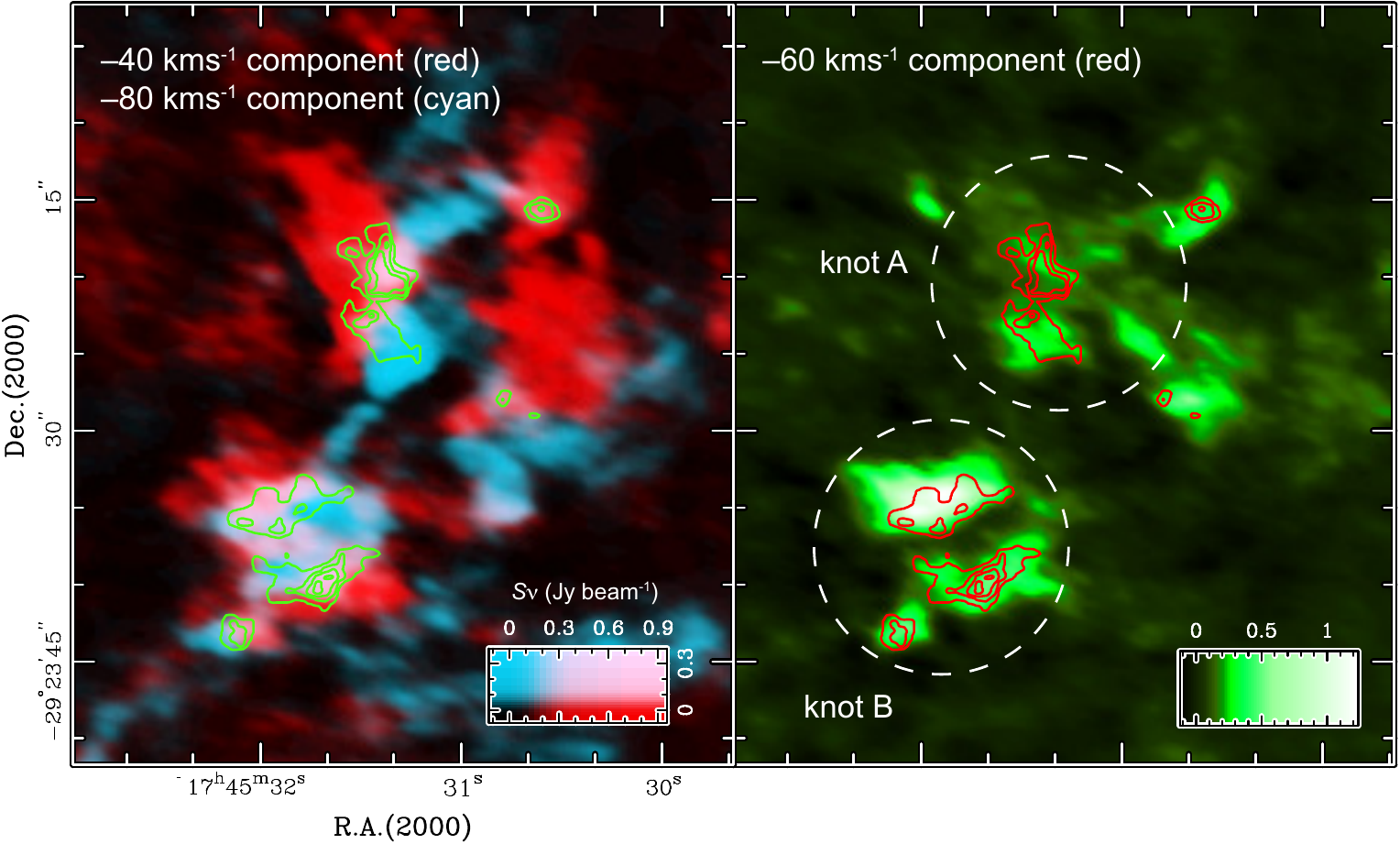}
\caption{(left panel) Composite color image of the HCN \JJ{3}{2}\ flux densities averaged over the velocity ranges [$-30$\ \kmps, $-50$\ \kmps] and [$-90$\ \kmps, $-70$\ \kmps] (in red and cyan, respectively) near knots A and B, with overlaid contours of \Dv\ at levels of 25, 30, 35, and 40 \kmps.  (right panel)  Map of the HCN \JJ{3}{2}\ flux density averaged over the velocity range [$-70$\ \kmps, $-50$\ \kmps], with the same contours of \Dv\ as those in the left panel.   The spatial distribution of the broad emission with $\Dv \ge 25\ \kmps$ is correlated well with the $-60$-\kmps\ component,  and appear almost exclusively in the boundary regions between the $-80$- and $-40$-\kmps\ components.\label{FIG_MULT_BROAD}}
\end{figure*}

\ifdraft
\begin{figure}[p!]
\else
\begin{figure}[h!]
\fi
\epsscale{1}
\plotone{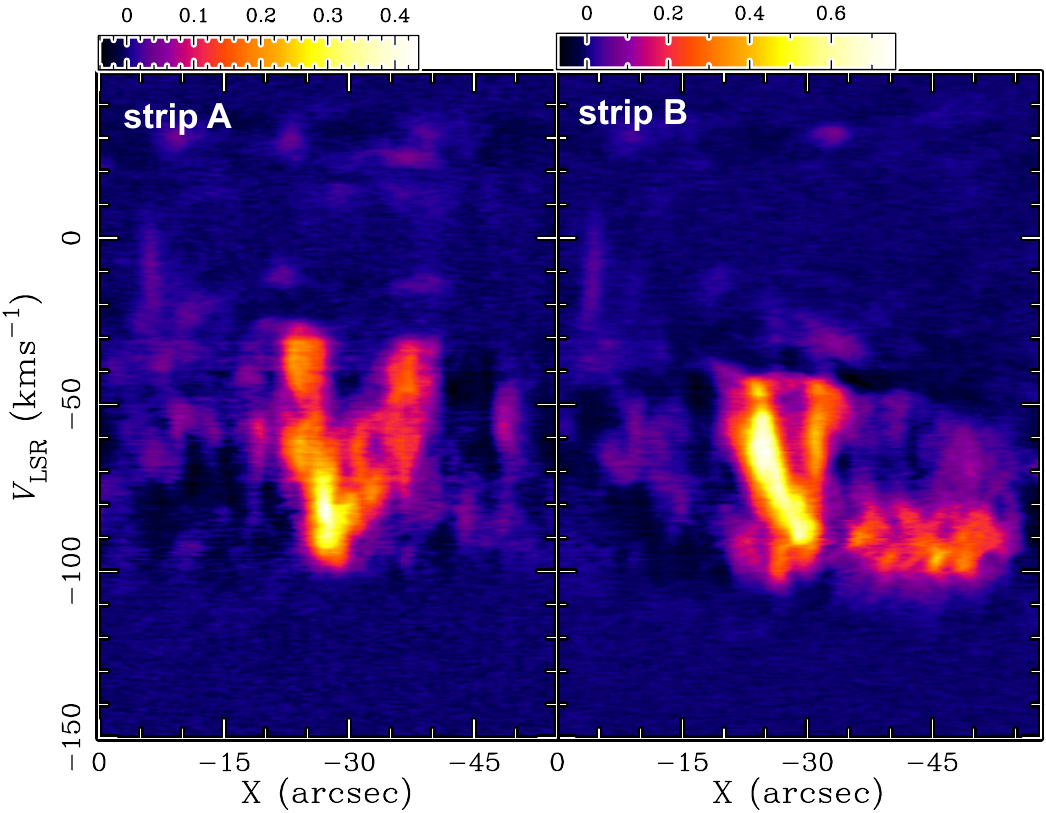}
\caption{PV diagram of the original HCN \JJ{3}{2}\ data along strips-A and -B across the $-80$-\kmps\ filament.   The strips are indicated by the rectangular regions in Figure \ref{FIG_DGRAM3}(b), where the $X$ axes are taken in the directions of the long sides of the rectangles. The spectra are averaged over the short sides.   The V-shaped patterns indicative of cloud--cloud interaction are clearly recognizable.\label{FIG_STRIPS}}
\end{figure}

\subsubsection{Moderately Broad-Line Emissions\label{SECTION_MOMENTANALYSIS}\label{SECTION_VELFIELD}}

\textcolor{red}{
Figure \ref{FIG_DGRAM3} shows the 2D maps of the zeroth moment, first moment (moment-0 and moment-1), and velocity width (\Dv), 
which are created from the reconstructed image of the main structure of the cloud.
}
The quantity \Dv\ is defined by the moment-0 values divided by the spectral-peak flux densities pixel-by-pixel.     
We also show the moment-1 map of the HCN map smoothed to a $20''$ resolution, i.e., the typical resolution of the previous single-dish observations \citep{Tanaka2014,Oka2016}\ in Figure \ref{FIG_DGRAM3}d.

The \Dv\ map (Figure \ref{FIG_DGRAM3}c) shows that individual spectra consisting \theObj\ are predominantly narrow; approximately 75\% of the pixels have \Dv\ narrower than 15\ \kmps, and the most frequent value is 10 \kmps.
These velocity widths are smaller than the variation in the moment-1 velocities over the entire cloud, $\sim 60\ \kmps$.
Pixels with \Dv\ larger than 25 \kmps\ are 5\%\ of the entire map, and their spatial distribution is concentrated to a few narrow areas near the $-80$-\kmps\ filament, \textcolor{red}{in particular to knots A and B}.
This confirms that the 90-\kmps\ width broad emission in the single-dish image consists of a superposition of multiple velocity components with narrow velocity widths.

\textcolor{red}{
Figure \ref{FIG_MULT_BROAD} is a close-up view of knots A and B, which presents the distributions of the three velocity components and the moderately broad emission with $\Dv > 25\ \kmps$. 
The figure shows that the $-40$- and $-80$-\kmps\ components are loosely anti-correlated with each other on the plane of the sky.  
In addition, the spatial distribution of the $-60$-\kmps\ component is correlated well with that of the moderately broad emission, and both extend along the narrow boundary regions between the $-80$- and $-40$-\kmps\ components.
This indicates that the moderately-broad emissions and the $-60$-\kmps\ component are the same physical entity, and they represent the transition region of the other two components in PPV space; 
this structure can be interpreted as that the line-broadening at knots A and B is caused by a physical interaction between $-80$- and $-40$-\kmps\ components.
}

\subsubsection{\textcolor{red}{V-Shaped Velocity Gradient Across the $-80$-\kmps\ Filament\label{SECTION_VGRAD}}}
Figure \ref{FIG_STRIPS} presents PV diagrams of the original HCN \JJ{3}{2}\ data for two cuts crossing the $-80$-\kmps\ filament at the locations of knots A and B.  
Along both cuts, the PV diagrams show characteristic V-shapes;  the $-80$-\kmps\ filament and the $-40$-\kmps\ component are smoothly bridged by a pair of bright emission features with steep velocity gradients.
These bridging emissions correspond to the moderately-broad $-60$-\kmps\ component identified with the dendrogram analysis.
\textcolor{red}{We note that this V-shape is not a PV pattern that appears only along these two particular cuts, but it persists over a wide spatial range along the filament.}
The moment-1 map (Figure \ref{FIG_DGRAM3}b) shows that the northern part of the filament from knot A to knot B is interposed by a pair of bands with $-60$--$-40$\ \kmps\ velocities that are running parallel to the filament;
this configuration is observed as the V-shaped pattern across cuts perpendicular to the filament.
Formation of this V-shaped PV pattern is readily explained by assuming physical interaction between the $-80$-\kmps\ filament and clouds at $-40$-\kmps.   
We will discuss this in more detail in a later section (\S\ref{DISCUSSION_CCC}).

\subsection{Heating Sources Inside or Near the $-80$-\kmps\ Filament\label{ROT_ANALY}}

\ifdraft
\begin{figure}[p!]
\else
\begin{figure}[t!]
\fi

\ifdraft
\epsscale{0.5}
\else
\epsscale{0.9}
\fi
\plotone{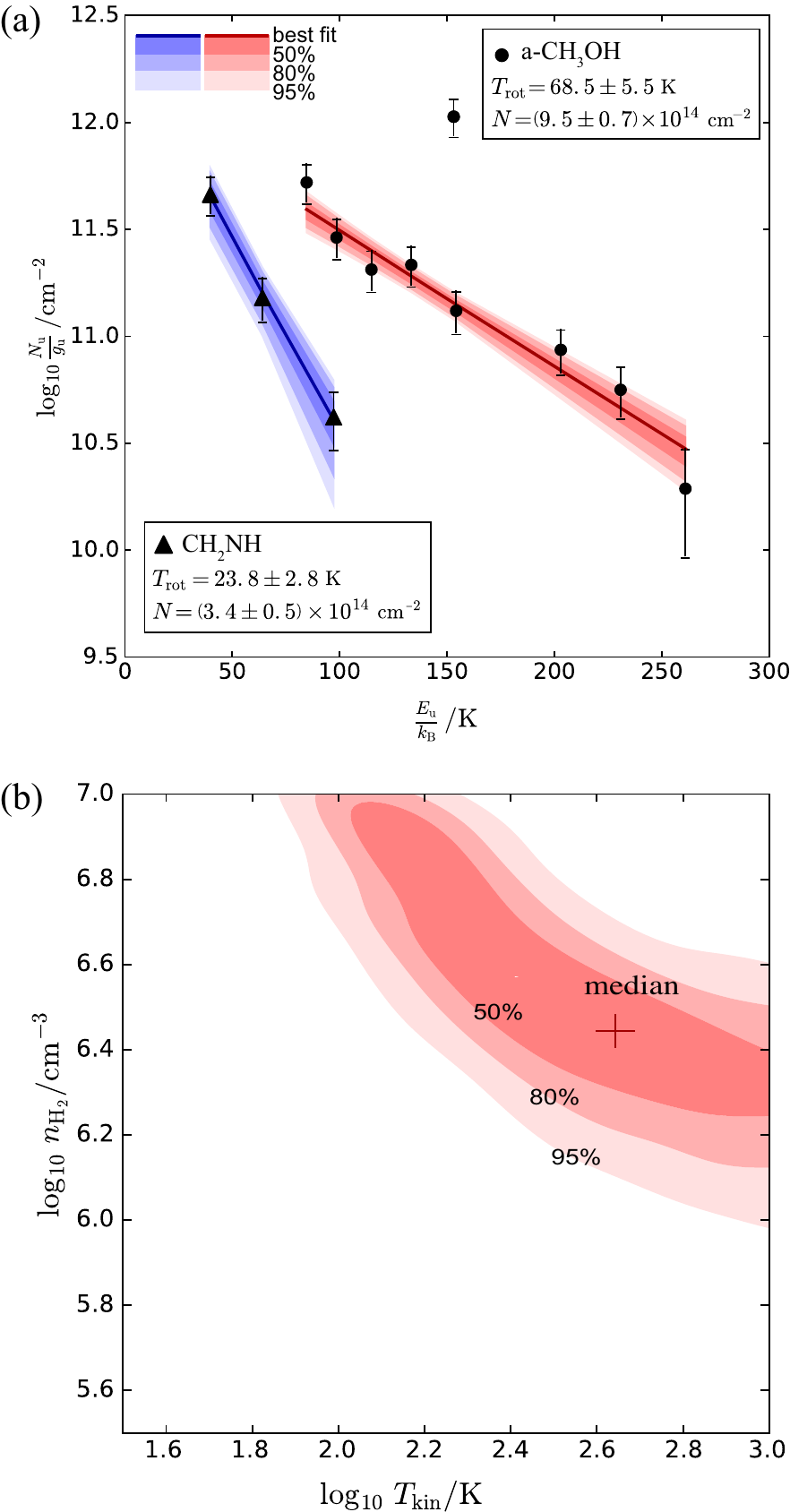}
\caption{(a) Rotational population diagram for \methanol\ and \methanimine\ in the $-80$-\kmps\ filament.  (b) Physical conditions in the filament calculated from a non-LTE analysis using the methanol lines.  Credible intervals obtained from a flat-prior Bayesian analysis are shown by the colored contours.\label{FIG_ROT}}
\end{figure}

We have investigated the physical conditions in the $-80$-\kmps\ filament by means of a population diagram.
Figure \ref{FIG_ROT}(a) shows the rotational population diagrams for the a-\methanol\ and \methanimine\  lines, constructed from the peak-intensity values of the spectra averaged over the five positions along the filament indicated in Figure \ref{FIG_FILAMENT}. 
We ignored the c-HCCCH line that blended with the 251.517 GHz \methanol\ line, assuming that the former is significantly weaker than the latter.
The error bars include both the r.m.s. noise measured for a blank-sky region in the data cube and conservatively assumed 20\% relative errors. 
The rate coefficients and partition functions are taken from the Leiden Atomic and Molecular Database \citep[LAMDA;][]{Schoier2005}, the Cologne database for molecular spectroscopy \citep[CDMA;][]{Muller2001}, and \cite{Motoki2014}.

Each of the a-\methanol\ and \methanimine\ level population diagrams is consistently fitted by a single-component, local thermodynamic equilibrium (LTE) curve except for the 250.507 GHz line of methanol, which is a predicted class-I maser line \citep{Voronkov2016}.
The diagram indicates two temperature components, corresponding to cold gas with a \methanimine\ rotational temperature $\Trot \sim 24\ $K and warm gas with a \methanol\ rotational temperature $\Trot \sim 70$ K.     
Although accurate values of \ncrit\ for the \methanimine\ lines are unknown, the similar A coefficients for the \methanol\ and \methanimine\ lines ($\sim 10^{-4}\ \mathrm{s}^{-1}$)  allow us to assume that they have similar values of \ncrit\ and hence that the different \Trot\ values reflect differences in the physical conditions of the emitting regions.

We performed a non-LTE analysis for the methanol lines using collisional excitation coefficients taken from LAMDA, where the gas kinetic temperature \Tkin, column density $N$, and hydrogen volume density \nHH, are taken as free parameters.   
Figure \ref{FIG_ROT}(b) shows the simultaneous credible intervals for \Tkin\ and \nHH, calculated from a Bayesian analysis assuming a prior function that is flat for $1 < \Tkin/\kelvin < 1000$ and $1 < \nHH/\pcc < 10^7$\ and zero otherwise.
Although the degeneracy between \Tkin\ and \nHH\ is not resolved, this result indicates that the filament has highly elevated \Tkin\ and/or \nHH\ compared with the physical conditions of standard Galactic center (GC) clouds,  for which \Tkin = $30$--$100$ \kelvin\ and $\nHH = 10^{4}\ \pcc$ \citep[e.g.][]{Martin2004,Nagai2007,Ginsburg2016,Arai2016}. 
In particular, this analysis indicates that $\nHH \gtrsim 10^6\ \pcc$, unless an unrealistically high \Tkin\ of $> 10^3$ K is assumed.
This result is also consistent with that from a non-LTE analysis using HCN lines \citep{Tanaka2014}. 

It is worthwhile noting that \Trot\ of 70 K for the warm component is among the highest of those measured for the CMZ clouds;  it is comparable to that measured toward the Sgr B2 hot cores.  In contrast, typical CMZ clouds have 10--20 K \methanol\ rotational temperatures \citep{Requena-Torres2006}.   
The co-existence of this exceptionally high-temperature gas and the cold component indicates the existence of some heating mechanism working near or inside the $-80$-\kmps\ filament.

The column densities of \methanimine\ and \methanol\ obtained from the non-LTE analysis are not particularly large compared with those in typical GC clouds or hot cores. 
We calculated the fractional abundances of \methanimine\ and \methanol\ to be $2\times10^{-8}$ and $1\times10^{-7}$, respectively, using hydrogen column densities obtained from the 265 GHz dust flux  assuming a dust temperature of 20 K and dust-emissivity index of 1.5.  These abundances are typical values for hot cores or Galactic center clouds \citep{Requena-Torres2006, Suzuki2016}.

\section{Discussion}

\subsection{Comparison with the Gravitational Kick Model\label{SECTION_ERROR}}
The observational bases of the gravitational-kick scenario presented by \cite{Oka2016,Oka2017} can be summarized as follows:
\begin{itemize}
\item The low-resolution images obtained with the NRO 45-m and ASTE 10-m telescopes show a slightly elongated elliptical shape.   \cite{Oka2016}\ detected a steep velocity gradient from $-110$ to $-20$\ \kmps\ across the major axis, and interpreted it as a beam-smeared image of a gas stream formed thorough IMBH--cloud interaction.  Although the stream is not fully resolved in the single-dish images, which have $\sim 20''$ angular resolutions, \cite{Oka2016} predicted its PV structure on the basis of numerical simulations.
\item \cite{Oka2017} identified a compact clump with an extremely broad emission of a $110$-\kmps\ width near the predicted position of the hypothesized IMBH, using the high-resolution image obtained with the ALMA observation.   The authors consider that this extremely-broad emission clump had been scattered by tidal interactions with the IMBH.   The PV position of the clump is approximately consistent with the trajectory of the stream predicted by \cite{Oka2016}.
\item \cite{Oka2017} argued that the compact continuum source MM1/CO$-0.4^*$\ is self-absorbed synchrotron emission from an IMBH, on the basis of the shallow spectral index of $\alpha = 1.18\pm0.65$ measured from the continuum fluxes at 230 GHz and 265 GHz.  The position of the source is consistent with the locus of the gravitational source in the model in \cite{Oka2016}.
\end{itemize}

In this section, we examine these arguments using our re-analyzed ALMA data.

\subsubsection{Stream}
First, we examine the significance of the steep velocity gradient detected in the single-dish images. 
The spatial width of the velocity gradient in \cite{Oka2016}\ is no greater than 2--3 times the resolution; 
\textcolor{red}{
in such under-resolved images, beam smearing effects can create an artificial gradient wherever multiple velocity components overlap.
In fact, the moment-1 map (Figure \ref{FIG_DGRAM3}b) shows the absence of a large-scale, monotonic velocity gradient on the entire-cloud scale; 
instead, the velocity field is dominated by very steep variations on $\sim 5''$ spatial scales created between the patches of $-40$-\kmps\ emissions and the widespread $-80$-\kmps\ emissions.
This indicates that the streamer-like velocity structure reported in \cite{Oka2016} is nonphysical, but instead is a product of the severe beam-smearing effect on the multicomponent velocity field.
We confirmed that the $20''$-resolution moment-1 map (Figure \ref{FIG_DGRAM3}d) exhibits an apparently smooth gradient that is consistent with that observed in the single-dish data.
}

Second, we examine the validity of the stream model directly by comparing it with the ALMA data.
The approximate shape of the predicted stream is overlaid on the methanol map (Figure \ref{FIG_FILAMENT}b). 
Obviously, a stream structure consistent with the model is not detected in any of the molecular or continuum images.
The most conspicuous stream-like structure in our data is the $-80$-\kmps\ filament, but its direction is perpendicular to that of the predicted stream. 
In addition, the filament lacks a systematic velocity gradient along its length that would be expected for a stream from a gravitationally kicked clump.
The velocity gradient is steepest across the width of the filament, and exhibits a V-shaped pattern rather than a monotonic increase or decrease. 
Such a gradient cannot be created through interaction with a point gravitational source outside the cloud;
\textcolor{red}{\cite{Lumine1986} have shown that the tidal deformation of a self-gravitating object passing a black hole is expressed by an affine transformation in the first approximation, which would not give rise to the observed prominent V-shaped pattern in the internal velocity field.}

\subsubsection{The 110-\kmps-Wide Emission}
The second feature, the extremely broad-line clump, is not detected in our re-analysis of the ALMA data used by \cite{Oka2017}.
Figure \ref{FIG_ERROR}(a) shows the PV diagram along cut C (Figure \ref{FIG_DGRAM3}b), which is the approximately same as cut b in \cite{Oka2017}.
Although the image presented in Figure 2 of \cite{Oka2017} has an extremely broad HCN emission extending from $-110\ \kmps$ to $0$\ \kmps, the spectra in the re-analyzed image are much narrower and completely lack emission for $\vlsr > -40\ \kmps$.
The measured velocity width \Dv\ is 40\ \kmps, which is larger than that of standard quiescent clouds but would not be peculiar enough to require an IMBH; 
similar or even broader linewidths are found at several other positions in knots A and B, as well as in high-resolution maps of other CMZ clouds \citep{Rathborne2014a,Higuchi2014,Tanaka2015}.

The inconsistency between \cite{Oka2017} and our present analysis is due to the difference in the data reduction processes.
The ALMA data contain two 12-m array observations of the HCN line, performed on July 8, 2013 and March 11, 2014.   
We excluded the former from our analysis because of insufficient data quality and a possible calibration failure, while both data were used by \cite{Oka2017}. 
\textcolor{red}{
We find that the reported extremely broad-lined emission can be created by merging the two datasets in the topocentric frequency frame (i.e., the coordinate frame where the telescope is at rest), rather than in the frame of the local standard of rest (LSR).
The topocentric frequency varies with dates due to the Doppler shift caused by the orbital motion of the Earth.
The topocentric frequency of the target differs by approximately $40$\ \kmps\ between July 8 and March 11;  the omission of correction for this drift causes severe blurring of the spectra. }
As the ALMA data uses the topocentric frame for frequency definition, a simple merge of multiple datasets using the {\tt CONCAT} task from the CASA package can introduce this error.
We confirmed that a PV diagram almost identical to that presented in \cite{Oka2017} is obtained from this incorrectly merged data, as shown in Figure \ref{FIG_ERROR}(b).

We note that the blurring effect does not only exaggerate the moderately broad-line $-60$-\kmps\ component;
it also dilutes emissions with line widths narrower than $40$\ \kmps, and masks many other PV structures in the cloud.
The narrow-lined $-40$-\kmps\ component, the $-80$-\kmps\ filament, and the V-shaped velocity gradients are mostly invisible in the erroneous data.

\subsubsection{Continuum Source MM1/CO$-0.4^*$}

\ifdraft
\begin{figure*}[p!]
\else
\begin{figure*}[t]
\fi
\epsscale{0.9}
\plotone{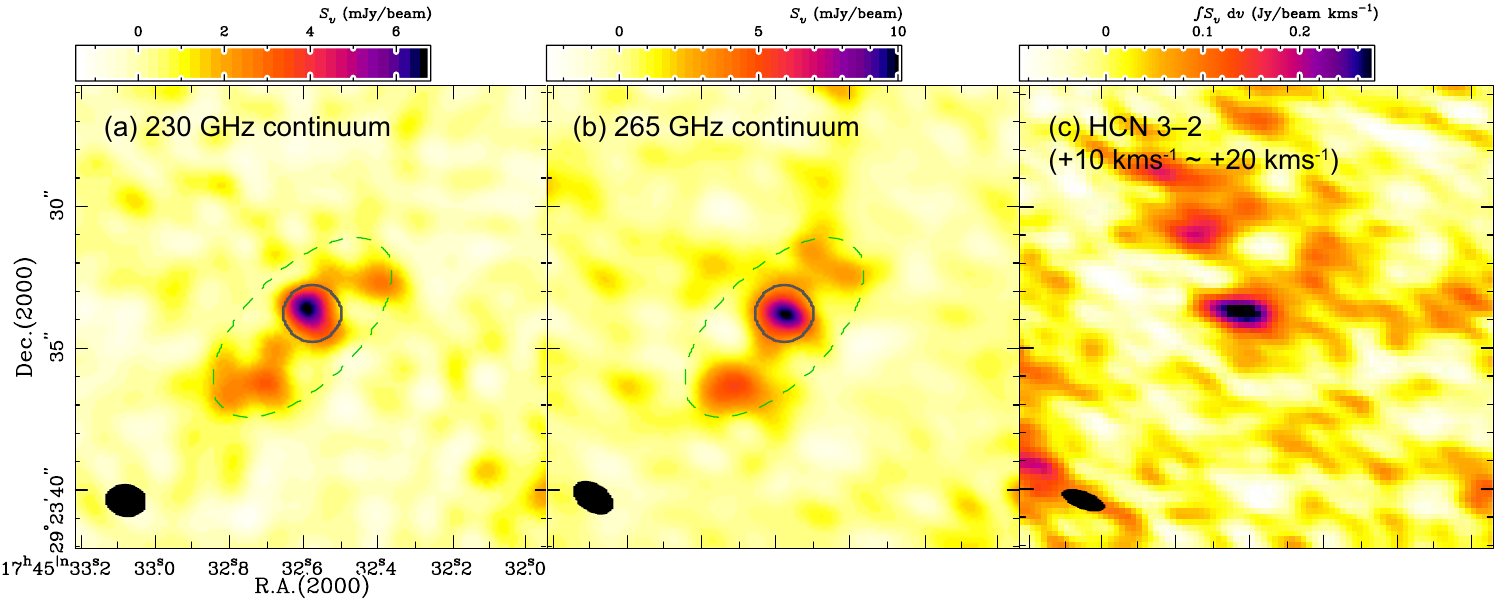}
\caption{Source MM1 in the 230 GHz  and 265 GHz continuum images (panels a and b, respectively), which are constructed from the $uv$ data within the uv-range of 20--1500 k$\lambda$.   The shapes of the synthesized beams are shown by the filled-ellipses.   The fluxes of MM1 and of the extended emission were measured for the $1''$-radius circle (the black solid line) and for an ellipse with semimajor and semiminor axes of $4''\times2''$ (the green broken line). Panel (c) shows the HCN \JJ{3}{2}\ flux densities integrated over the \vlsr\ range from $10\ \kmps$ to $20\ \kmps$.\label{FIG_MM1}}
\end{figure*}

The shallow spectral index $\alpha \sim 1$ measured for the narrow band of the ALMA data \citep{Oka2017}\ contradicts the reported non-detection at 34.25 GHz above the 3-$\sigma$ upper limit of 0.285 mJy \citep{Ravi2017}, which requires $\alpha\gtrsim 2$.
The reason for this inconsistency is likely to be the omission of beam-size corrections by \cite{Oka2017}.
Although the synthesized beam areas of their continuum images are 1.74 arcsec$^2$ and 0.74 arcsec$^2$  at 230 GHz and 265 GHz, respectively, they measured the fluxes by summing the pixel values inside the beam of the 265 GHz data for both frequencies;
hence, the effective beam area for the flux measurement at 230 GHz is approximately twice that at 265 GHz. 
This leads to an underestimation of $\alpha$ since MM1 is not an ideal point source but includes non-negligible background emission. 

We obtained a steeper spectral index from the same data as used by \cite{Oka2017} by applying a beam size correction.
Figure \ref{FIG_MM1} shows the 230 GHz and 265 GHz continuum images of MM1, which we constructed using the $uv$ data within the spatial frequency range of 20--1500\ k$\lambda$ to correct for the differences in $uv$ coverage.
The synthesized beam sizes are $1.44''\times1.16''$ and $1.57''\times1.00''$ at 230 GHz and 265 GHz, respectively.
We further smoothed the images so that they both have an effective beam size of $2''$ FWHM, since the center positions  of the source at the two frequencies differ by approximately half the original beam.
The corrected flux densities are $8.6\pm0.2$ mJy and $13.2\pm0.4$ mJy at 230 GHz and 265 GHz, respectively, which gives a spectral index of $\alpha = 3.0\pm0.3$. 
Although our measurement is contaminated by the background emission, 
the net spectral index of MM1 should not differ significantly from that obtained here, since measurement of the extended emission (indicated by the broken-lined ellipse in Figure \ref{FIG_MM1}) gives a similar value of $\alpha = 2.5\pm 0.4$.

Although it is difficult to identify the nature of MM1 from the ALMA data alone, the steep value of $\alpha \sim 3$ is typical for thermal dust emission.  
This supports the argument of \cite{Ravi2017} that the source could be emission from warm dust in circumstellar material, whereas 
interpretations as optically thin synchrotron emission from a relativistic jet or wind \citep{Ravi2017} are not favored.
The hypothesis of synchrotron emission from an IMBH \citep{Oka2017} might be marginally consistent with our result, as self-absorbed synchrotron emission can have values of $\alpha$ up to 2.5, but
 assuming such an exotic object would not be reasonable when the value of $\alpha$ is common for thermal dust emission.

The thermal origin for MM1 is also suggested by the detection of a molecular line counterpart. 
Figure \ref{FIG_MM1}(c) presents the HCN \JJ{3}{2}\ integrated intensity for the \vlsr\ range from $10\ \kmps$ to $20\ \kmps$, which shows a compact source at the position of MM1.  As other velocity channels lack intensity peaks at this position, this 15-\kmps\ source is most likely a  molecular-line counterpart to MM1.   
The systemic LSR velocity of 15 \kmps\ suggests that the source is associated with the foreground cloud G$359.62{-0.24}$, for which the distance has been measured to be $3.1^{+2.2}_{-0.9}$ kpc through parallax measurements using a water maser source \citep{Iwata2017}.

\bigskip
To conclude, the signature of a gravitational interaction with an IMBH is not confirmed in the re-analyzed data.
The features previously reported as evidence for the IMBH scenario are most likely due to severe beam smearing in the single-dish images, blurring of the spectra introduced by errors in the data reduction, and insufficient correction for the difference in the beam sizes between the two measurement frequency bands.

\ifdraft
\begin{figure}[p!]
\else
\begin{figure}[t!]
\fi
\epsscale{1}
\plotone{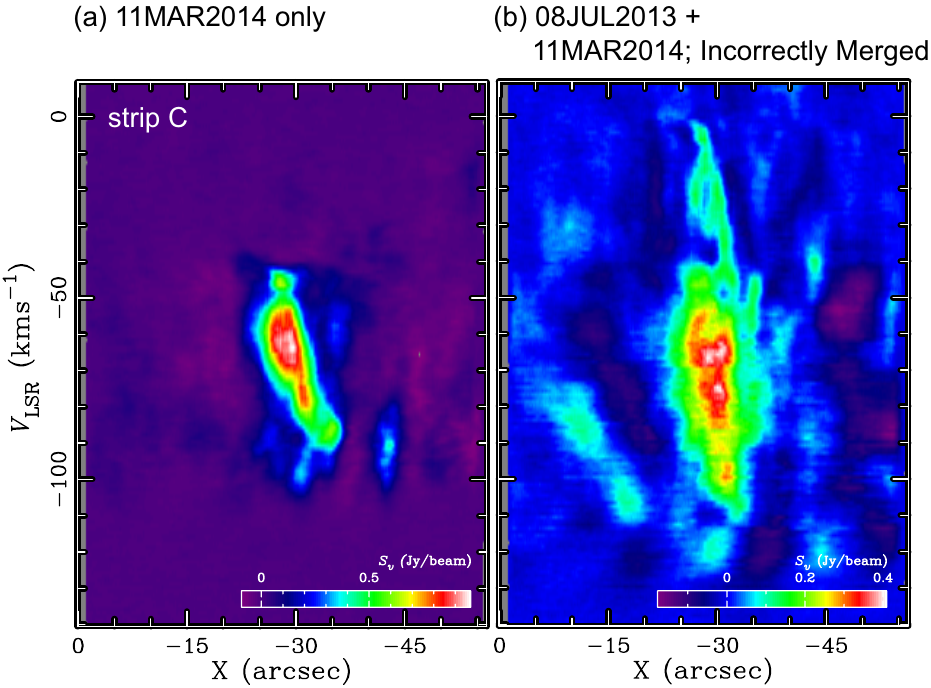}
\caption{ (a) The HCN \JJ{3}{2}\ position-velocity diagram along cut C of Figure \ref{FIG_DGRAM3}(b), which is approximately the same as cut b in Figures 1 and 2 of \cite{Oka2017}.  (b) Same as (a), but for erroneous data, in which the {\tt 08JUL2013} and \texttt{11MAR2014} datasets are merged without correcting for their different topocentric frequencies.   The LSR velocities are calculated using the latter date.  The $110\ \kmps$-wide extremely broad emission is only reproduced in the erroneous data.\label{FIG_ERROR}}
\end{figure}

\subsection{Cloud--Cloud Collision Hypothesis\label{DISCUSSION_CCC}}

\ifdraft
\begin{figure*}[p!]
\else
\begin{figure*}[t!]
\fi
\epsscale{1.}
\plotone{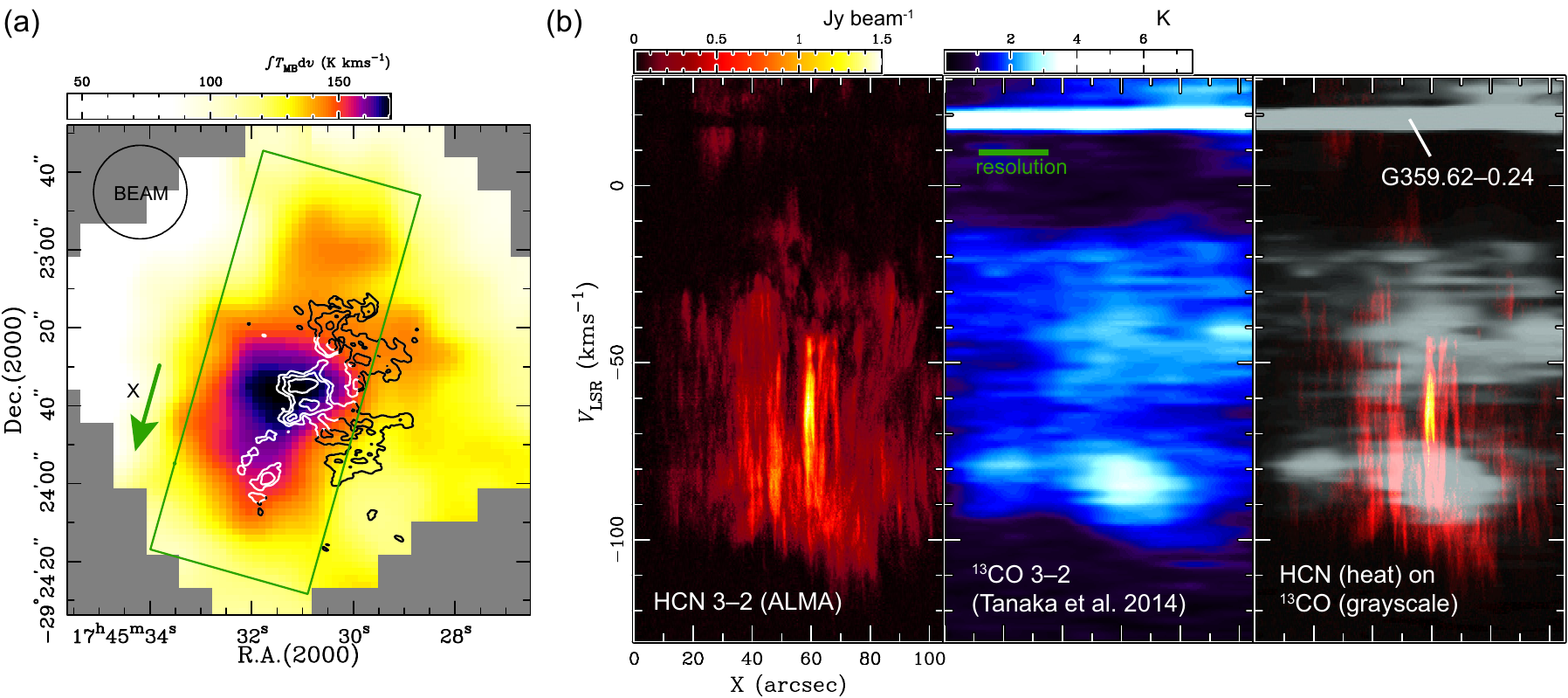}
\caption{(a) \COt\ \JJ{3}{2}\ integrated intensity map for the $-110$ to $0$\ \kmps\ velocity range obtained with the ASTE 10-m telescope \citep{Tanaka2014}, with contours of the ALMA \JJ{3}{2}\ integrated flux density drawn at 5, 10, 20, 50, and 100 Jy\,$\mathrm{beam^{-1}}\cdot\kmps$.  (b) PV diagrams for the ALMA HCN \JJ{3}{2}\ and  ASTE \COt\ \JJ{3}{2}\ data (left and middle panels, respectively) for the rectangular field in panel (a), where the $X$ axis is in the direction of the long side of the rectangle, and the spectra are averaged in the direction of the short side.   \textcolor{red}{The bright emission band in the 15--$20\ \kmps$ velocity channels of the \COt\ diagram is the foreground infrared dark cloud G359.62{$-0.24$}.}  The HCN and \COt\ images are overlaid in the rightmost panel.\label{FIG_SINGLEDISH_13CO}}
\end{figure*}

\ifdraft
\begin{figure}[p!]
\else
\begin{figure}[t!]
\fi
\epsscale{1}
\plotone{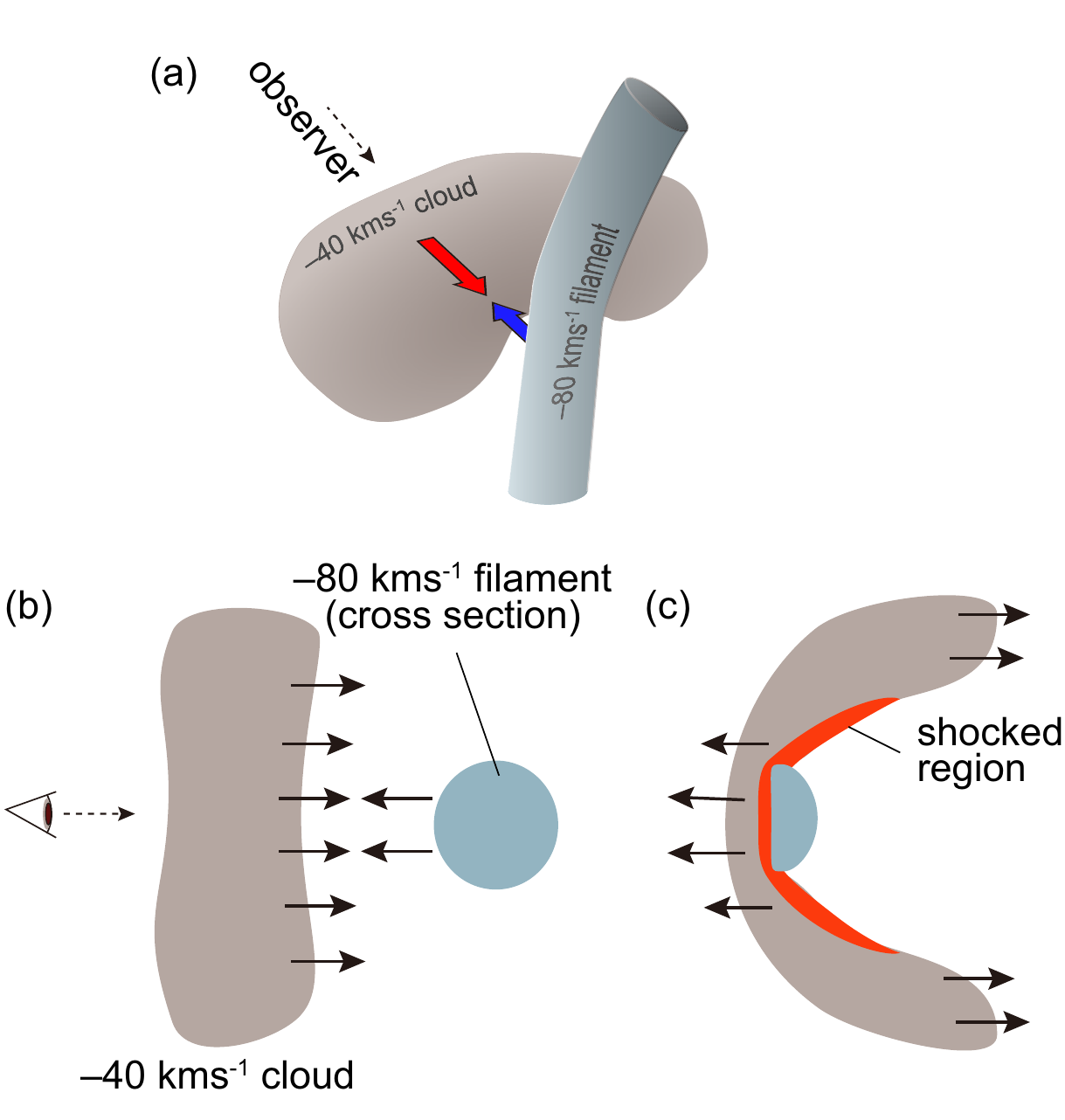}
\caption{Schematic illustration of a collision between $-40$-\kmps\ cloud and the $-80$-\kmps\ filament, based on theoretical calculations of a CCC system with two dissimilar clouds \citep{Habe1992,Anathpindika2010,Takahira2014}.  (a) The $-80$-\kmps\ filament and the $-40$-\kmps\ cloud immediately before the collision. (b) Same as (a) but in a cross section view.  (c) The collision phase.  The central portion of the $-40$-\kmps\ cloud is pushed forward by the $-80$-\kmps\ filament, while the outer part remains undisturbed.\label{FIG_MANGA}}
\end{figure}

The dendrogram analysis has shown that the characteristic broad-lined emission from \theObj\ does not originate from  one coherent kinematical feature, but instead is a superposition of multiple velocity components. 
The $-40$- and $-80$-\kmps\ components have narrow velocity widths, $\Dv = 10$ \kmps, whereas the $-60$-\kmps\ feature consists of a moderately broad emission with $\Dv\ = 25$--$50\ \kmps$.
The moment analysis described in Sections \ref{SECTION_VELFIELD} and \ref{SECTION_VGRAD} has revealed that these components are smoothly connected at knots A and B on the $-80$-\kmps\ filament in PPV space;
the small-scale distributions of the $-80$-\kmps\ and $-40$-\kmps\ components inside the knots exhibit a spatial anti-correlation on the plane of the sky, but these two primary velocity components are bridged by the moderately broad emissions that extend along the boundary between them.
Such a configuration, consisting of two narrow-line component and a broad emission bridging them, is a signpost of a CCC system \citep{Haworth2015a,Torii2017a}.
In this CCC hypothesis, Knots A and B are interpreted as the positions at which the $-80$-\kmps\ filament collides with the $-40$-\kmps\ cloud. 

The bridging emission can be seen more clearly by comparing the ALMA HCN data with the single-dish \COt\ \JJ{3}{2}\ map \citep{Tanaka2014}, in which spatially extended or low-density components unobservable with ALMA are visible.
Figure \ref{FIG_SINGLEDISH_13CO}\ shows the position-velocity diagrams of the single-dish \COt\ \JJ{3}{2} and the ALMA HCN \JJ{3}{2} data along a cut approximately parallel to the $-80$-\kmps\ filament, as indicated in Figure \ref{FIG_SINGLEDISH_13CO}(a). 
The \COt\ map has counterparts to the $-40$- and $-80$-\kmps\ components of the ALMA data, but lacks the $-60$-\kmps\ component.   
The \COt\ line is generally optically thin and primarily traces the mass distribution of the gas, whereas the HCN \JJ{3}{2}\ is sensitive to gas with a low column density, but higher temperature and gas volume density are required to excite it. 
Therefore, the absence of  bridging emission in the \COt\ map indicates that \theObj\ is separated into two distinct dynamical components represented by the $-40$- and $-80$-\kmps\ components.
Superposition of the single-dish and ALMA data presented in Figure \ref{FIG_SINGLEDISH_13CO}(b) shows clearly that the broad-lined HCN emission bridges the velocity gaps between the two primary components. 
This bridging emission is minor in mass, but has a highly elevated temperature and density;
it can be interpreted as a thin layer of turbulent gas created by the collision of the two clouds.

This CCC hypothesis is strengthened by the detection of the V-shaped PV pattern across the width of the $-80$-\kmps\ filament.  
Theoretical studies of CCC systems of two non-identical clouds \citep{Habe1992,Anathpindika2010,Takahira2014} predict the formation of bow shocks inside the larger cloud.
If the same model is applicable to a collision between a narrow filament and an extended cloud, the shocked region should have a V-shaped PV pattern when the collision interface is observed in the face-on geometry. 
We show a schematic illustration of such a collision of the $-80$-\kmps\ filament and the $-40$-\kmps\ cloud in Figure \ref{FIG_MANGA}.   
The central portion of the $-40$-\kmps\ cloud is dragged along by the plunging $-80$-\kmps\ filament and the bow shock, while the motion of the outer part of the $-40$-\kmps\ cloud remains less affected.  
As the intrinsic velocity widths in the pre-collision clouds are smaller than the line-of-sight collision velocity,  
this gas kinematics is expected to be observed as a V-shaped velocity pattern when viewed along the collision axis \citep{Takahira2014,Haworth2015a,Torii2017a}.
Similar models have been proposed for the formation of half cavities in the Brick cloud, the Sgr B2 complex, and the $50$-\kmps\ cloud \citep{Higuchi2014,Tsuboi2015a,Tsuboi2015b}.

The CCC model also explains the mechanism responsible for  heating the $-80$-\kmps\ filament to the observed high temperature. 
By balancing the heating produced by dissipation of the velocity dispersion of 10\ \kmps\ in the knots against the radiative cooling by gas and dust \citep{Ao2013}, we obtain 220 K for the gas temperature, where we used  velocity gradient ${\rm d}v/{\rm d}r = 2\times10^2\ \kmps\pc^{-1}$,  filament width $L = 0.1\ \pc$, and $\nHH = 10^6\ \pcc$  for the calculation.  
This temperature is within the parameter range obtained from the non-LTE analysis of the multi-transition methanol lines.
\textcolor{red}{
\subsection{Other Hypotheses}
\cite{Ravi2017} proposed a CCC hypothesis that is different from ours, in which the infrared dark cloud G359.62${-0.24}$ is assumed to have fallen from the Galactic halo and collided with \theObj.
However, PV patterns indicative of physical interactions between G359.62${-0.24}$ and \theObj\ are not identified in the ALMA HCN \JJ{3}{2}\ data (Figure \ref{FIG1}a) nor in the single-dish \COt\ data (Figure \ref{FIG_SINGLEDISH_13CO}).
The parallax measurement toward a water maser source in G359.62${-0.24}$ \citep{Iwata2017}\ suggests that the source is most likely in the foreground spiral arm region.
}

\textcolor{red}{
Another hypothesis by \cite{Yalinewich2017} assuming SN--cloud interactions as the origin of the HVCCs does not conflict with our CCC scenario.
As we have shown in a previous paper \citep{Tanaka2014},  \theObj\ is located on the rim of an expanding-shell-like structure of dense molecular gas with diffuse radio continuum emission, which could be understood as an SN-driven shell.
This expanding shell is a plausible candidate of the trigger of the CCC event at \theObj.
}

\subsection{Cloud--Cloud Collisions in the Galactic Center Region}

The large-scale CO surveys toward the CMZ have detected approximately $100$ HVCCs and HVCC-like broad-line features \citep{Oka2012,Tokuyama2017}.  
\textcolor{red}{
Signatures of CCC similar to those found in \theObj\ have been detected in another archetypical HVCC, CO${-0.30}{-0.07}$ \citep{Tanaka2015}.
If a significant fraction of the HVCCs in the CMZ are CCC sites similar to those archetypical HVCCs,
it would indicate a high CCC frequency in the CMZ, since HVCCs are rather short-lived features.
}
By naively assuming that every HVCC indicates one CCC event, the CCC frequency can be estimated from the number of HVCCs, and the duration of the HVCC phase, $\tau_{\rm HVCC}$.  
On ignoring  radiative disruption induced by star formation activity \citep{Haworth2015},  
we obtain $\tau_{\rm HVCC} \sim r/\sigma_v = 7\times10^4\ \yr$, where $r = 1\ \pc$ and $\sigma_v = 15\ \kmps$ are typical size and velocity dispersion for an HVCC, respectively.
Given that a typical cloud mass in the CMZ is $\sim 10^4\ \Msol$ \citep{Miyazaki2000} and that the total cloud mass of the CMZ is $3\times10^7\ \Msol$\citep{Molinari2011}, the CCC frequency is estimated to be 3 per cloud in one orbital period \citep[6 Myr;][]{Sofue2013}.
The actual CCC frequency could be somewhat higher, since whether a CCC system is detectable as a broad emission feature may depend on the viewing angle; 
\cite{Haworth2015a} estimate the detection efficiency to be 20--30\%\ due to this effect.
In total, the CCC rate is roughly estimated to be an order of $1$--$10$ per cloud per orbital period, if CCC is the primary origin of the HVCCs.
This is in good agreement with the theoretically predicted CCC frequency: a few to 15 times per orbital period \citep{Tan2000,Tasker2009,Dobbs2015}.
Although these theoretical calculations are not necessarily tuned for the Galactic center environment \citep{Dobbs2015}, it is unlikely that the CCC frequency in the CMZ is lower than in the Galactic disk.  
The high volume filling factor of dense gas, fast orbital velocities, and frequent SN explosions may further increase the CCC frequency.

CCC-triggered star formation (SF) is among the popular theories for the formation of the massive clusters in the CMZ \citep{Hasegawa1994,Stolte2008}.
Our analysis has shown that \nHH\ in the post-shock region is $\gtrsim 10^6\ \pcc$, which is approximately two orders of magnitude higher than the average \nHH\ in  CMZ clouds.   
This indicates that CCC is actually able to compress the gas to  densities typical of star-forming cores during the dynamical time of a collision.  
However, direct evidence has not been obtained for current SF in \theObj; 
\textcolor{red}{the compact continuum source MM1 is likely to be in the foreground Galactic disk cloud and not associated with \theObj.} 
Radio continuum sources indicative of \ion{H}{2}\ regions are not detected in existing centimeter-wavelength images \citep[e.g.][]{Liszt1995}.
Although the bright methanol spots on the $-80$-\kmps\ filament are similar  to hot cores,  their high temperature and rich abundances of complex organic molecules (COMs) could also be an immediate result of a shock passage.    Indeed, the fractional methanol abundance of $10^{-7}$ is approximately equal to the average value for non-star-forming clouds in the CMZ \citep{Requena-Torres2006}, where the high COM abundance is thought to be maintained by mechanical sputtering of dust mantles, not by thermal desorption from hot cores.

One explanation for the lack of SF signatures may be that the cloud is at an early phase in a CCC-triggered SF process, in which SF has not yet been activated.  
Alternatively, it is possible that the collision is non-productive.    CCCs  may stabilize clouds by enhancing turbulent pressure, or they may even destroy the clouds if the ram pressure of the collision exceeds the gravitational binding pressure \citep{Habe1992,Tan2000,Dobbs2011,Johnston2014,Kruijssen2014,Rathborne2014a}.
The general absence of active SF in the HVCCs and the CCC candidate regions --- except for the Sgr B2 complex \citep[e.g.][]{Oka2012,Rathborne2014a,Higuchi2014,Tanaka2015}\ --- may indicate that CCCs in the CMZ are predominantly in such a non-productive regime.  
Considering that every CMZ cloud may experience more than one CCC during its lifetime, non-productive CCCs may be a non-negligible mechanism for suppressing the star formation efficiency in the GC region \citep{Kauffmann2017}.

\section{Summary}

We have reanalyzed the ALMA archival data for the HVCC \theObj, and have explored the origin of the broad-profiled molecular line emission of the cloud.
In particular, through a careful investigation of the velocity field, we have examined the recently proposed hypothesis that the cloud has been gravitationally kicked by an IMBH. 
Our main conclusions are summarized below:

\begin{itemize}
\item The high-resolution ALMA image of the HCN \JJ{3}{2}\ line shows that the broad-profiled emission detected with single-dish observations consists of a superposition of multiple velocity components: two narrow-line ($\Dv\sim 10\ \kmps$ FWZI) components at \vlsr = $-80$ and $-40$\ \kmps, and a moderately broad ($\Dv\sim 25$--$50\ \kmps$) emission component at $-60$\ \kmps.    

\item A major fraction of the cloud mass is concentrated in a straight filament with an approximately constant velocity of $-80$\ \kmps.  This $-80$-\kmps\ filament intersects the $-40$-\kmps\ clouds at two positions on the plane of the sky (knots A and B), where the velocity width is the largest in the entire cloud.  The $-80$- and $-40$-\kmps\ emissions are smoothly bridged in the PPV space at the knots, forming V-shaped PV patterns across in the direction crossing the filament.  The moderately broad-line emission of the $-60$-\kmps\ component almost exclusively appears in these features with 'V'-letter-shaped PV patterns.

\item The coherent stream structure predicted by the IMBH hypothesis of \cite{Oka2016} is not detected.   The steep velocity gradient obtained in the single-dish images, which forms the basis for their model, is found instead to be a beam-smeared feature resulting from the superposition of multiple velocity components.     In addition, the gravitationally-kicked cloud feature with a $110$-\kmps\ velocity width reported by \cite{Oka2017} is not reproduced in our re-analysis of the same ALMA data.  Instead, this feature is highly likely to be an artifact due to the blurring of spectra caused by the omission of temporal drift of the Doppler-shifted topocentric sky frequencies \textcolor{red}{due to the orbital motion of the Earth}.  

\item The spectral index of the IMBH candidate source MM1/CO${-0.40}{-0.22}^*$ is found to be $\alpha = 3.0\pm0.4$ from the fluxes at 230 GHz and 265 GHz, which is consistent with a recently reported value of $\alpha \gtrsim 2$ measured with 34.25 GHz observations \citep{Ravi2017}.  The shallow $\alpha$ of $1.18\pm0.65$ reported by \cite{Oka2017} is most likely due to larger contamination by background emission in the 230 GHz flux measurement.   As a molecular line counterpart is detected at \vlsr = 15\ \kmps, this source is likely to be a thermal dust emission from the foreground cloud G${359.62}{-0.24}$ in the Galactic disk.

\item The PV structure of \theObj\ agrees well with that predicted for a collision between two dissimilar-sized clouds;  the broad emission component at $-60$ \kmps\ corresponds to the bridging emission that represents the shocked gas layer created by a collision between the $-80$- and $-40$-\kmps\ components.  The portion of the $-40$-\kmps\ cloud that interacts with the narrow $-80$-\kmps\ filament is compressed and dragged in the direction of the relative motion of the filament, creating the V-shaped PV pattern across the filament width.   

\item None of the evidence previously reported for the IMBH hypothesis is confirmed by this re-analysis.  Instead, we argue that a CCC scenario is the most plausible hypothesis for the origin of \theObj.    If the $\sim 100$ HVCCs detected in wide-field survey maps are predominantly sites of CCCs, we obtain a high CCC frequency of order  1--10 per cloud per orbital period.   

\item We have measured the  volume density of the post-shock gas to be $\gtrsim 10^6\ \pcc$ from a non-LTE analysis of multi-transition methanol lines.  This indicates that a CCC shock can increase the gas density by two orders of magnitude during the dynamical time scale of a collision.   Direct evidence for on-going star formation in \theObj\ has not been obtained.   The general absence of the star formation signatures in HVCCs may indicate that they are still at an early phase of CCC-triggered star formation, or that CCCs are non-productive in the Galactic center region.   If the latter is the case, the frequent CCC may contribute to suppress the star formation efficiency in the CMZ.
\end{itemize}

\acknowledgements
The author is grateful to the anonymous referee, whose suggestions and comments helped much to improve the paper. 
This work was supported by JSPS KAKENHI Grant Number 16K17666.
\textcolor{red}{
This paper makes use of the following ALMA data: ADS/JAO.ALMA\#2012.1.00940.S. ALMA is a partnership of ESO (representing its member states), NSF (USA) and NINS (Japan), together with NRC (Canada), MOST and ASIAA (Taiwan), and KASI (Republic of Korea), in cooperation with the Republic of Chile. The Joint ALMA Observatory is operated by ESO, AUI/NRAO and NAOJ.
}

\appendix
\section{Dendrogram Analysis\label{appendix:dendrogram}}
We employed the dendrogram analysis \citep{Rosolowsky2008} to extract the primary PPV structure of \theObj\ from the highly complicated HCN \JJ{3}{2}\ data cube (\S\ref{SECTION_DENDRO}).
Clumps were identified in the data cube that was smoothed into $2.5\ \kmps$ velocity bins and digitized in units of $3\sigma$ intensities ($\sigma = 13$ mJy). 
The tree diagram presented in Figure \ref{FIG_DGRAMTREE}\ shows the hierarchical structure of the identified clumps;  
the nodes in the diagram represent clumps that are defined as isolated closed isosurfaces at given threshold levels, and the vertical branches connecting the nodes mean that the lower-level clumps are split into multiple clumps at the higher threshold levels.  
Single-node trees growing directly from the ground \citep[``sprouts'';][]{Lee2014} are omitted in the diagram.

The thick-lined tree in the diagram represents the main body of \theObj.
We identified six main branches in this cloud tree, which are colored other than in black in the diagram. 
The spatial distributions of these six main branches are superposed on the moment-0 map presented in Figure \ref{FIG_DGRAM3}(a).   Branches 1, 4, and 5 correspond to the southern, northern, and middle parts of the $-80$-\kmps\ filament.   
Knots A and B are included in Branches 5 and 4, respectively.
An image of the basic structure of the cloud was reconstructed by adding the six main branches and their common trunk nodes (indicated by the thick black lines in Figure \ref{FIG_DGRAMTREE});  
this procedure removed from the original image minor leaf clumps and emissions not related to the main body of the cloud. 

\ifdraft
\begin{figure}[p!]
\else
\begin{figure}[t!]
\fi
\ifdraft
\epsscale{.5}
\else
\epsscale{.5}
\fi
\plotone{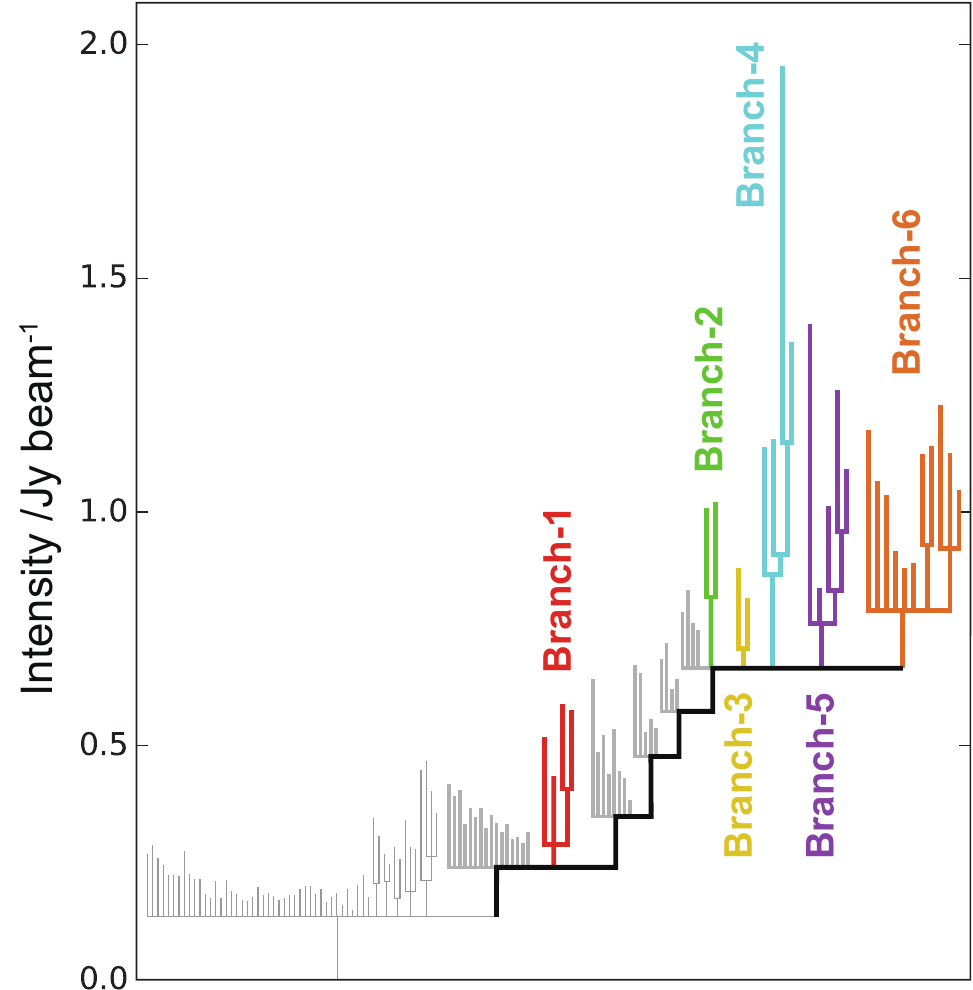}
\caption{Dendrogram for the clumps identified in the HCN \JJ{3}{2}\ PPV map.   Leaf clumps growing directly from the ground level (``sprouts'') are omitted.  The thick-lined trees represent the \theObj\ main body.  The six main branches are colored other than black.\label{FIG_DGRAMTREE}}
\end{figure}

\bibliographystyle{apj}
\bibliography{mendeley,local}

\newcommand{\noop}[1]{}
\begin{thebibliography}{}
\expandafter\ifx\csname natexlab\endcsname\relax\def\natexlab#1{#1}\fi

\bibitem[{Anathpindika(2010)}]{Anathpindika2010}
Anathpindika, S.~V. 2010, \mnras, 405, 1431

\bibitem[{Ao {et~al.}(2013)Ao, Henkel, Menten, Requena-Torres, Stanke,
  Mauersberger, Aalto, M{\"{u}}hle, \& Mangum}]{Ao2013}
Ao, Y., Henkel, C., Menten, K.~M., {et~al.} 2013, \aap, 550, A135

\bibitem[{Arai {et~al.}(2016)Arai, Nagai, Fujita, Nakai, Seta, Yamauchi,
  Kaneko, Hagiwara, Mamyoda, Miyamoto, Horie, Ishii, Koide, Ogino, Maruyama,
  Hirai, Oshiro, Nagai, Akiyama, Konakawa, Nonogawa, Salak, Terabe, Nihonmatsu,
  \& Funahashi}]{Arai2016}
Arai, H., Nagai, M., Fujita, S., {et~al.} 2016, \pasj, 68, 2

\bibitem[{Bally {et~al.}(2010)Bally, Aguirre, Battersby, Bradley, Cyganowski,
  Dowell, Drosback, Dunham, {Evans II}, Ginsburg, Glenn, Harvey, Mills,
  Merello, Rosolowsky, Schlingman, Shirley, Stringfellow, Walawender, \&
  Williams}]{Bally2010}
Bally, J., Aguirre, J., Battersby, C., {et~al.} 2010, \apj, 721, 137

\bibitem[{Dobbs {et~al.}(2011)Dobbs, Burkert, \& Pringle}]{Dobbs2011}
Dobbs, C.~L., Burkert, A., \& Pringle, J.~E. 2011, \mnras, 413, 2935

\bibitem[{Dobbs {et~al.}(2015)Dobbs, Pringle, \& Duarte-Cabral}]{Dobbs2015}
Dobbs, C.~L., Pringle, J.~E., \& Duarte-Cabral, A. 2015, \mnras, 446,
  3608

\bibitem[{Ginsburg {et~al.}(2016)Ginsburg, Henkel, Ao, Riquelme, Kau, Pillai,
  Mills, Requena-Torres, Immer, Testi, Ott, Bally, Battersby, Darling, Aalto,
  Stanke, Kendrew, Kruijssen, Longmore, Dale, Guesten, \&
  Menten}]{Ginsburg2016}
Ginsburg, A., Henkel, C., Ao, Y., {et~al.} 2016, \aap, 50, 1

\bibitem[{Habe \& Ohta(1992)}]{Habe1992}
Habe, A., \& Ohta, K. 1992, \pasj, 44, 203

\bibitem[{Hasegawa {et~al.}(1994)Hasegawa, Sato, Whiteoak, \&
  Miyawaki}]{Hasegawa1994}
Hasegawa, T., Sato, F., Whiteoak, J.~B., \& Miyawaki, R. 1994, \apjl,
  429, L77

\bibitem[{Haworth {et~al.}(2015{\natexlab{a}})Haworth, Shima, Tasker, Fukui,
  Torii, Dale, Takahira, \& Habe}]{Haworth2015}
Haworth, T.~J., Shima, K., Tasker, E.~J., {et~al.} 2015{\natexlab{a}},
  \mnras, 454, 1634

\bibitem[{Haworth {et~al.}(2015{\natexlab{b}})Haworth, Tasker, Fukui, Torii,
  Dale, Shima, Takahira, Habe, \& Hasegawa}]{Haworth2015a}
Haworth, T.~J., Tasker, E.~J., Fukui, Y., {et~al.} 2015{\natexlab{b}},
  \mnras, 450, 10

\bibitem[{Higuchi {et~al.}(2014)Higuchi, Chibueze, Habe, Takahira, \&
  Takano}]{Higuchi2014}
Higuchi, A.~E., Chibueze, J.~O., Habe, A., Takahira, K., \& Takano, S. 2014,
  \aj, 147, 141

\bibitem[{Iwata {et~al.}(2017)Iwata, Kato, Sakai, \& Oka}]{Iwata2017}
Iwata, Y., Kato, H., Sakai, D., \& Oka, T. 2017, \apj, 840, 18

\bibitem[{Johnston {et~al.}(2014)Johnston, Beuther, Linz, Schmiedeke, Ragan, \&
  Henning}]{Johnston2014}
Johnston, K.~G., Beuther, H., Linz, H., {et~al.} 2014, Astronomy {\&}
  Astrophysics, 568, A56

\bibitem[{Kauffmann {et~al.}(2017)Kauffmann, Pillai, Zhang, Menten, Goldsmith,
  Lu, Guzm{\'{a}}n, \& Schmiedeke}]{Kauffmann2017}
Kauffmann, J., Pillai, T., Zhang, Q., {et~al.} 2017, \aap, 603, 89

\bibitem[{Kruijssen {et~al.}(2014)Kruijssen, Longmore, Elmegreen, Murray,
  Bally, Testi, \& Kennicutt}]{Kruijssen2014}
Kruijssen, J. M.~D., Longmore, S.~N., Elmegreen, B.~G., {et~al.} 2014,
  \mnras, 440, 3370

\bibitem[{Lee {et~al.}(2014)Lee, Fern, Storm, Looney, Mundy, Arce, Ostriker,
  Shirley, Segura-cox, Teuben, Rosolowsky, Kwon, Kauffmann, Tobin, Plunkett,
  Pound, \& Salter}]{Lee2014}
Lee, K.~I., Fern, M., Storm, S., {et~al.} 2014, \apj, 76, 76

\bibitem[{Liszt \& Spiker(1995)}]{Liszt1995}
Liszt, H.~S., \& Spiker, R.~W. 1995, \apjs, 98, 259

\bibitem[{Luminet \& Carter(1986)}]{Lumine1986}
Luminet, J.-P., \& Carter, B. 1986, \apjs, 61, 219

\bibitem[{Martin {et~al.}(2004)Martin, Walsh, Xiao, Lane, Walker, \&
  Stark}]{Martin2004}
Martin, C.~L., Walsh, W.~M., Xiao, K., {et~al.} 2004, \apjs, 150, 239

\bibitem[{Miyazaki \& Tsuboi(2000)}]{Miyazaki2000}
Miyazaki, A., \& Tsuboi, M. 2000, \apj, 536, 357

\bibitem[{Molinari {et~al.}(2011)Molinari, Bally, Noriega-Crespo,
  Compi{\`{e}}gne, Bernard, Paradis, Martin, Testi, Barlow, Moore, Plume,
  Swinyard, Zavagno, Calzoletti, {Di Giorgio}, Elia, Faustini, Natoli,
  Pestalozzi, Pezzuto, Piacentini, Polenta, Polychroni, Schisano, Traficante,
  Veneziani, Battersby, Burton, Carey, Fukui, Li, Lord, Morgan, Motte,
  Schuller, Stringfellow, Tan, Thompson, Ward-Thompson, White, \&
  Umana}]{Molinari2011}
Molinari, S., Bally, J., Noriega-Crespo, a., {et~al.} 2011, \apj, 735,
  L33

\bibitem[{Motoki {et~al.}(2014)Motoki, Isobe, Ozeki, \& Kobayashi}]{Motoki2014}
Motoki, Y., Isobe, F., Ozeki, H., \& Kobayashi, K. 2014, \aap, 28, 1

\bibitem[{Muller {et~al.}(2001)Muller, Thorwirth, Roth, \&
  Winnewisser}]{Muller2001}
Muller, H. S.~P., Thorwirth, S., Roth, D.~A., \& Winnewisser, G. 2001,
  \aap, 52, L49

\bibitem[{Nagai {et~al.}(2007)Nagai, Tanaka, Kamegai, \& Oka}]{Nagai2007}
Nagai, M., Tanaka, K., Kamegai, K., \& Oka, T. 2007, \pasj, 59, 25

\bibitem[{Oka {et~al.}(2016)Oka, Mizuno, Miura, \& Takekawa}]{Oka2016}
Oka, T., Mizuno, R., Miura, K., \& Takekawa, S. 2016, \apjl, 816, L7

\bibitem[{Oka {et~al.}(2012)Oka, Onodera, Nagai, Tanaka, Matsumura, \&
  Kamegai}]{Oka2012}
Oka, T., Onodera, Y., Nagai, M., {et~al.} 2012, \apjs, 201, 14

\bibitem[{Oka {et~al.}(2017)Oka, Tsujimoto, Iwata, Nomura, \&
  Takekawa}]{Oka2017}
Oka, T., Tsujimoto, S., Iwata, Y., Nomura, M., \& Takekawa, S. 2017, Nature
  Astronomy, 1, 709

\bibitem[{Rathborne {et~al.}(2014)Rathborne, Longmore, Jackson, Foster,
  Contreras, Garay, Testi, Alves, Bally, Bastian, Kruijssen, \&
  Bressert}]{Rathborne2014a}
Rathborne, J., Longmore, S.~N., Jackson, J.~M., {et~al.} 2014, \apj,
  786, 140

\bibitem[{Ravi {et~al.}(2017)Ravi, Vedantham, \& Phinney}]{Ravi2017}
Ravi, V., Vedantham, H., \& Phinney, E.~S. 2017, arXiv,
  1710.03813v1

\bibitem[{Requena-Torres {et~al.}(2006)Requena-Torres, Mart{\'{i}}n-Pintado,
  Rord{\'{i}}guez-Franco, Mart{\'{i}}n, Rodr{\'{i}}guez-Fern{\'{a}}ndez, \&
  de~Vicente}]{Requena-Torres2006}
Requena-Torres, M.~A., Mart{\'{i}}n-Pintado, J., Rord{\'{i}}guez-Franco, N.~J.,
  {et~al.} 2006, \aap, 455, 971

\bibitem[{Rosolowsky {et~al.}(2008)Rosolowsky, Pineda, Kauffmann, \&
  Goodman}]{Rosolowsky2008}
Rosolowsky, E., Pineda, J.~E., Kauffmann, J., \& Goodman, A.~A. 2008,
  \apj, 1338

\bibitem[{Sch{\"{o}}ier {et~al.}(2005)Sch{\"{o}}ier, Tak, Dishoeck, \&
  Black}]{Schoier2005}
Sch{\"{o}}ier, F.~L., Tak, F. F. S. V.~D., Dishoeck, E. F.~V., \& Black, J.~H.
  2005, \aap, 432, 369

\bibitem[{Sofue(2013)}]{Sofue2013}
Sofue, Y. 2013, \pasj, 65, 118

\bibitem[{Stolte {et~al.}(2008)Stolte, Ghez, Morris, Lu, Brandner, \&
  Matthews}]{Stolte2008}
Stolte, A., Ghez, A.~M., Morris, M.~R., {et~al.} 2008, \apj, 675, 1278

\bibitem[{Suzuki {et~al.}(2016)Suzuki, Ohishi, Hirota, Saito, Majumdar, \&
  Wakelam}]{Suzuki2016}
Suzuki, T., Ohishi, M., Hirota, T., {et~al.} 2016, \apj, 825, 1

\bibitem[{Takahira {et~al.}(2014)Takahira, Tasker, \& Habe}]{Takahira2014}
Takahira, K., Tasker, E.~J., \& Habe, A. 2014, \apj, 792, 63

\bibitem[{Tan(2000)}]{Tan2000}
Tan, J.~C. 2000, \apj, 536, 173

\bibitem[{Tanaka {et~al.}(2015)Tanaka, Nagai, Kamegai, \& Oka}]{Tanaka2015}
Tanaka, K., Nagai, M., Kamegai, K., \& Oka, T. 2015, \apj, 806, 130

\bibitem[{Tanaka {et~al.}(2014)Tanaka, Oka, Matsumura, Nagai, \&
  Kamegai}]{Tanaka2014}
Tanaka, K., Oka, T., Matsumura, S., Nagai, M., \& Kamegai, K. 2014,
  \apj, 783, 62

\bibitem[{Tanaka {et~al.}(2009)Tanaka, Oka, Nagai, \& Kamegai}]{Tanaka2009}
Tanaka, K., Oka, T., Nagai, M., \& Kamegai, K. 2009, \pasj, 61, 461

\bibitem[{Tasker \& Tan(2009)}]{Tasker2009}
Tasker, E.~J., \& Tan, J.~C. 2009, \apj, 700, 358

\bibitem[{Tokuyama {et~al.}(2017)Tokuyama, Oka, Takekawa, Yamada, Iwata, \&
  Tsujimoto}]{Tokuyama2017}
Tokuyama, S., Oka, T., Takekawa, S., {et~al.} 2017, Proceedings of the
  International Astronomical Union, 11, 154

\bibitem[{Torii {et~al.}(2017)Torii, Hattori, Matsuo, Fujita, Nishimura, Kohno,
  Kuriki, Tsuda, Minamidani, Umemoto, Kuno, Yoshiike, Ohama, Tachihara, Fukui,
  Shima, Habe, \& Haworth}]{Torii2017a}
Torii, K., Hattori, Y., Matsuo, M., {et~al.} 2017, \pasj, 835, 142

\bibitem[{Tsuboi {et~al.}(2015a)Tsuboi, Miyazaki, \& Uehara}]{Tsuboi2015a}
Tsuboi, M., Miyazaki, A., \& Uehara, K. 2015, \pasj, 67, 1

\bibitem[{Tsuboi {et~al.}(2015b)Tsuboi, Miyazaki, \& Uehara}]{Tsuboi2015b}
Tsuboi, M., Myyazaki, A., \& Uehara, K. 2015, \pasj, 67, 90

\bibitem[{Voronkov {et~al.}(2016)Voronkov, Caswell, Ellingsen, \&
  Breen}]{Voronkov2016}
Voronkov, M.~A., Caswell, J.~L., Ellingsen, S.~P., \& Breen, S.~L. 2016, Cosmic
  Masers - from OH to H0, Proceedings of the International Astronomical Union,
  IAU Symposium,, 287, 433

\bibitem[{Yalinewich \& Beniamini(2017)}]{Yalinewich2017}
Yalinewich, A., \& Beniamini, P. 2017, arXiv, 1709.05738v1

\end{thebibliography}

\end{document}